\documentclass[reprint,nofootinbib,amsmath,amssymb,aps,prd,superscriptaddress]{revtex4-2}

\pdfoutput=1

\usepackage{graphicx}
\usepackage{dcolumn}
\usepackage{bm}
\usepackage{hyperref}

\usepackage{lipsum}

\usepackage{graphicx}	% Including figure files
\usepackage{amsmath}	% Advanced maths commands
\usepackage{amssymb}	% Extra maths symbols

\usepackage[usenames,dvipsnames]{color}
\usepackage[normalem]{ulem}

\usepackage{enumitem}

\usepackage{tabularx}
\newcolumntype{L}{>{\arraybackslash}X} %This produces justified text
\usepackage{multirow}

\newcommand{\tabitemize}[1]{%
\begin{minipage}[t]{\linewidth}
\begin{itemize}[nosep,left=0pt]
#1
\end{itemize}
\end{minipage}}

\newcommand{\dif}{\mathrm{d}}
\newcommand{\const}{\mathrm{const.}}

\newcommand{\e}{\mathrm{e}}

\DeclareMathOperator{\sign}{sign}

\DeclareMathOperator*{\arcsinh}{arcsinh}
\DeclareMathOperator*{\arccosh}{arccosh}

\DeclareMathOperator*{\Si}{Si}
\DeclareMathOperator*{\Ci}{Ci}
\DeclareMathOperator*{\Shi}{Shi}
\DeclareMathOperator*{\Chi}{Chi}

\newcommand{\matter}{\mathrm{matter}}

\newcommand{\stress}{\LC{T}}

\newcommand\LC[1]{{#1}{}}
\newcommand\LCG{\LC{\Gamma}}

\newcommand\LCbox{\LC{\square}}
\newcommand\LCR{\LC{R}}

\newcommand\LCEin{\LC{G}}
\newcommand\LCKre{\LC{\mathcal{K}}}

\usepackage{calrsfs}

\DeclareMathAlphabet{\pazocal}{OMS}{zplm}{m}{n}
\usepackage[scr=boondox]{mathalfa}

\newcommand{\bigo}{\mathcal{O}}

\usepackage{xcolor}

\usepackage{amsthm}
\newtheorem{result}{Result}
\newtheorem*{corollary*}{Corollary}

\newcommand{\myskip}{\hspace{9pt}}

\begin{document}

\title{Physical non-viability of a wide class of $f(R)$ models and their constant-curvature solutions}

\author{Adri\'{a}n Casado-Turri\'{o}n}
    \email{adricasa@ucm.es}
    \affiliation{Departamento de F\'{i}sica Te\'{o}rica and Instituto IPARCOS,  Universidad Complutense, 28040 Madrid, Spain}
\author{\'{A}lvaro de la Cruz-Dombriz}
    \email{alvaro.dombriz@usal.es}
    \affiliation{ Departamento de F\'{i}sica Fundamental, Universidad de Salamanca, 37008 Salamanca, Spain}
    \affiliation{Cosmology and Gravity Group, Department of Mathematics and Applied Mathematics, University of Cape Town, Rondebosch 7700, Cape Town, South Africa}
\author{Antonio Dobado}
    \email{dobado@fis.ucm.es}
    \affiliation{Departamento de F\'{i}sica Te\'{o}rica and Instituto IPARCOS,  Universidad Complutense, 28040 Madrid, Spain}

\date{\today}

\begin{abstract}
Constant-curvature solutions lie at the very core of gravitational physics, with Schwarzschild and (Anti)-de Sitter being two of the most paradigmatic examples. Although such kind of solutions are very well-known in General Relativity, that is not the case for theories of gravity beyond the Einsteinian paradigm. In this article, we provide a systematic overview on $f(R)$ models allowing for constant-curvature solutions, as well as of the constant-curvature solutions themselves. We conclude that the vast majority of these $f(R)$ models suffer, in general, from several shortcomings rendering their viability extremely limited, when not ruled out by physical evidence. Among these deficiencies are instabilities (including previously unforeseen strong-coupling problems) and issues limiting the predictive power of the models. Furthermore, we will also show that most $f(R)$-exclusive constant-curvature solutions also exhibit a variety of unphysical properties.

\end{abstract}

\maketitle

\section{Introduction}

When modelling physical systems, it is often necessary to resort to simplifying assumptions, either to make the equations describing the problem more tractable or to gain further insight on the relevant physics. It seems obvious that such assumptions should be physically well-motivated and consistent with both experiments and the theoretical framework being employed. For example, within the General Relativity (GR) framework, the explanation of a variety of cosmological observations relies on the crucial assumption that the Universe is \textit{approximately} homogeneous and isotropic at sufficiently large scales.

However, despite its success in describing most gravitational phenomena, GR still suffers from several shortcomings, such as its inability to describe dark energy without introducing a new, \textit{ad hoc} fluid in the theory. This and other weaknesses can be solved by generalising GR, for instance by postulating that the gravitational Lagrangian is given by a function $f(\LCR)$ of the Ricci scalar $\LCR$, instead of just $\LCR$ (as in Einsteinian gravity). Though simple, the $f(\LCR)$ ansatz turns out to comply with the basic consistency requirements outlined above in most circumstances. Except in some pathological examples, $f(\LCR)$ models are mathematically consistent, can be reduced to standard GR in the appropriate limit, and yield predictions which are in accordance with observations. Indeed, appropriate choices of function $f$ lead to a correct description of both early- and late-universe physics, such as inflation and the aforementioned dark-energy-dominated epoch. $f(\LCR)$ models have also found applications in stellar physics, with some of them being compatible with neutron-star and gravitational-wave observations \cite{Olmo:2019flu,Feola:2019zqg,Astashenok:2017dpo,Astashenok:2020qds,Ananda:2007xh,Capozziello:2008fn,Bouhmadi-Lopez:2012piq}.

Nonetheless, outside of the highly-symmetric cosmological scenarios, it is often very complicated to solve the fourth-order equations of metric $f(\LCR)$ gravity without making further simplifying assumptions. Currently, one of the most popular choices is to find solutions with constant scalar curvature $R$. Indeed, some of the most well-known solutions of GR have constant curvature, such as the Schwarzschild or Vaidya spacetimes, their generalisations including a cosmological constant, or the FLRW spacetime sourced by radiation. It has been known for a long time that many $f(\LCR)$ gravity models host only the same vacuum constant-curvature solutions as GR \cite{delaCruz-Dombriz:2009pzc}. However, as will be further detailed in Section \ref{section:constant-curvature solutions}, there is a particular set of $f(\LCR)$ models satisfying some additional assumptions \cite{Nzioki:2009av,Calza:2018ohl} for which \textit{any} metric with a given constant Ricci scalar is a solution of said $f(\LCR)$ model. Because of this, in what follows we shall refer to these special $f(\LCR)$ models admitting all spacetimes with $\LCR=\LCR_0=\const$ as \textit{$\LCR_0$-degenerate $f(\LCR)$ models}. In order to find new constant-curvature solutions of $\LCR_0$-degenerate $f(\LCR)$ models which are not present in GR, one merely needs to solve equation $\LCR=\LCR_0=\const$ for some particular metric ansatz and initial conditions.

For these reasons, it is almost immediate to obtain novel, $f(\LCR)$-exclusive constant-curvature solutions that provide an answer to virtually \textit{every} open problem in gravitational physics. For example, $f(\LCR)$-exclusive constant-curvature solutions describing wormholes made out of pure vacuum (and thus complying with the standard energy conditions), exotic black holes, or even spacetimes giving rise to the observed rotation curves of galaxies without introducing dark matter have been reported in the literature \cite{Duplessis:2015xva,Calza:2018ohl,Hendi:2020umk,Bertipagani:2020awe}. The approach has furthermore been generalised to other modified gravity theories, such as $f(Q)$, where a similar situation occurs \cite{Calza:2022mwt}.

In the present work, we shall show that the vast majority of $\LCR_0$-degenerate $f(\LCR)$ models, as well as their constant-curvature solutions themselves, are most often pathological in nature, and thus not physically viable. In particular, we will prove that this special class of $f(\LCR)$ models only propagates one scalar degree of freedom at linear level (in contrast with generic $f(\LCR)$ models and GR), thus being incompatible with gravitational-wave observations. Furthermore, as we shall discuss, $\LCR_0$-degenerate $f(\LCR)$ models apparently lack predictability, because of the aforementioned infinite degeneracy of their constant-curvature solutions. Finally, to make matters worse, we will also show that most of the novel constant-curvature solutions are unstable and host a number of unphysical properties, such as regions in which the metric signature changes abruptly, naked curvature singularities, and more.

The contents of this work will be organised as follows. First, in Section \ref{section:constant-curvature solutions}, we shall briefly review the conditions $f(\LCR)$ models must satisfy so as to harbour any metric with a given constant Ricci scalar. Second, the pathological character of the $f(\LCR)$ models within this special class will be discussed in Section \ref{Section: Pathologies of the models}. More precisely, the fact that the linearised spectrum of said models contains at most one massless scalar field only will be proven therein. Next, in Section \ref{Section: Stability of solutions}, we shall assess the stability of $f(\LCR)$-exclusive constant-curvature solutions under small perturbations of their Ricci scalar. Finally, Section \ref{introducing the solutions} will be devoted to a characterisation of several classes of novel constant-curvature solutions, some of which have not been yet reported in the literature, as far as we are concerned. We shall then conclude that most of the solutions analysed in the latter section display a variety of unphysical properties.

The busy reader is encouraged to focus on Section \ref{section:constant-curvature solutions}, containing our precise definitions of \textit{constant-curvature solutions} and \textit{$\LCR_0$-degenerate $f(\LCR)$ models}; Results \ref{Result: Non-propagation of gravity} and \ref{Result: Non-propagation of the scalaron}, containing our most relevant findings regarding the existence of strong-coupling instabilities in $\LCR_0$-degenerate $f(\LCR)$ models; Results \ref{Result: R0 instability} and \ref{Result: R0=0 metastability}, concerning the stability of constant-curvature solutions within $\LCR_0$-degenerate $f(\LCR)$ models; Table \ref{tab:solutions}, summarising all the pathological traits displayed by the $f(\LCR)$-exclusive solutions discussed herein; and, finally, the conclusions and final discussions collected in Section \ref{Section: Conclusions}.

Before proceeding with the results of our investigations, let us enumerate, for the sake of clarity, the various notational conventions to be followed hereafter. Our sign choice shall be the one denoted as $(-,-,-)$ by Misner, Thorne and Wheeler \cite{Misner:1973prb}: the metric signature will be $(+,-,-,-)$, the Riemann and Ricci tensors are defined as $\LCR^{\rho}{}_{\sigma\mu\nu}\equiv-2(\partial_{[\mu|}\LCG^{\rho}{}_{\sigma|\nu]}+\LCG^{\rho}{}_{\lambda[\mu|}\LCG^{\lambda}{}_{\sigma|\nu]})$ and $\LCR_{\mu\nu}\equiv\LCR^{\rho}{}_{\mu\rho\nu}$, respectively, and the Einstein field equations read $\LCEin_{\mu\nu}=-\kappa\stress_{\mu\nu}$, with $\kappa\equiv 8\pi G$ ($c=1$) and $\stress_{\mu\nu}\equiv+(2/\sqrt{-g})\,\delta S_\matter/\delta g^{\mu\nu}$, where $S_\matter$ is the matter action sourcing the gravitational sector. Moreover, as widely known, the total action of metric $f(\LCR)$ gravity coupled to matter reads
\begin{equation} \label{f(R) action}
    S\,=\,\dfrac{1}{2\kappa}\int\dif^4 x\,\sqrt{-g}\,f(\LCR)+S_\matter,
\end{equation}
whose associated equations of motion are
\begin{equation} \label{f(R) EOM}
    f'(\LCR)\LCR_{\mu\nu}-\dfrac{f(\LCR)}{2}g_{\mu\nu}+\mathcal{D}_{\mu\nu}f'(\LCR)=-\kappa\,\stress_{\mu\nu},
\end{equation}
where $\mathcal{D}_{\mu\nu}\equiv\nabla_\mu\nabla_\nu-g_{\mu\nu}\LCbox$ and $n$ primes right after any function denote the $n$-th derivative of said function with respect to its argument. For instance, $f'(\LCR)\equiv\dif f(\LCR)/\dif\LCR$ and $A''(r)\equiv\dif^2 A(r)/\dif r^2$.

\section{Constant-curvature vacuum solutions of $f(\LCR)$ gravity} \label{section:constant-curvature solutions}

Throughout this article, we shall define a \textit{constant-curvature spacetime} as the one represented by a metric whose Ricci scalar is constant, i.e.
\begin{equation}
    \LCR=\const\equiv\LCR_0.
\end{equation}
When the equations of motion of $f(\LCR)$ gravity \eqref{f(R) EOM} are evaluated in vacuum---i.e.~$\stress_{\mu\nu}=0$---and constant scalar curvature $\LCR_0$ solutions are seeked, the last two terms on the left-hand side of \eqref{f(R) EOM} vanish. Thus, the equations of motion reduce to
\begin{equation} \label{reduced f(R) EOM}
    f'(\LCR_0)\LCR_{\mu\nu}=\dfrac{f(\LCR_0)}{2}g_{\mu\nu}.
\end{equation}
Taking the trace of \eqref{reduced f(R) EOM}, one finds that, in vacuum, such a constant-curvature solution satisfies
\begin{equation}
\label{reduced f(R) EOM trace}
    f'(\LCR_0)\LCR_0=2f(\LCR_0).
\end{equation}
Thus, in the event that vacuum solutions with constant-curvature $\LCR_0$ are present in a given $f(\LCR)$ model, equations \eqref{f(R) EOM}, \eqref{reduced f(R) EOM} and \eqref{reduced f(R) EOM trace} hold simultaneously, giving rise to the following scenarios:
\begin{itemize}
    \item If $\LCR_0=0$, equation \eqref{reduced f(R) EOM trace} necessarily implies that $f(0)=0$. As a result, \eqref{reduced f(R) EOM} then entails that either $f'(0)=0$ or $\LCR_{\mu\nu}=0$. This means that an $f(\LCR)$ model satisfying $f(0)=0$ always admits the same $\LCR_0=0$ solutions as GR (for which $\LCR_{\mu\nu}=0$). If, in addition, $f'(0)=0$, then the full equations of motion \eqref{reduced f(R) EOM}---or, equivalently, \eqref{f(R) EOM}---are satisfied automatically, and the theory admits \textit{any} metric having $\LCR_0=0$ as a solution, even if said vanishing-curvature metrics are not solutions of GR. Notice that these novel, $f(\LCR)$-exclusive solutions would coexist with those of GR in $f(\LCR)$ models satisfying $f'(0)=0$.
    \item If $\LCR_0\neq 0$, there are two possibilities within this scenario.

        On the one hand, for $f(\LCR)$ models satisfying $f(\LCR_0)\neq 0$, equation \eqref{reduced f(R) EOM} necessarily implies that $f'(\LCR_0)\neq0$ (since $\LCR_0\neq 0$). Thus, equations \eqref{f(R) EOM} and \eqref{reduced f(R) EOM} turn into $\LCR_{\mu\nu}=(\LCR_0/4)g_{\mu\nu}$, and only the constant-curvature solutions of $\text{GR}+\Lambda$ ---with $\Lambda=\LCR_0/4=f(\LCR_0)/2f'(\LCR_0)$--- solve the equations of motion of the $f(\LCR)$ model under consideration.  
        
        On the other hand, for $f(\LCR)$ models satisfying $f(\LCR_0)=0$, equation \eqref{reduced f(R) EOM} forces $f'(\LCR_0)=0$. As such, equations \eqref{f(R) EOM} and \eqref{reduced f(R) EOM} are trivially satisfied, and \textit{any} metric with constant scalar curvature $\LCR_0$ is a solution of the $f(\LCR)$ model. In particular, the constant-curvature solutions of $\text{GR}+\Lambda$ (with $\Lambda=\LCR_0/4$) would also be solutions of this particular set of $f(\LCR)$ models. Thus, in these models, the novel, $f(\LCR)$-exclusive constant-curvature solutions (which do not satisfy the usual Einstein equations in the presence of a cosmological constant) would coexist with the constant-curvature solutions of Einsteinian gravity (with an appropriate cosmological constant).
\end{itemize}
In summary, \textit{any} metric with vanishing Ricci scalar trivially solves the vacuum equations of motion of \textit{all} $f(\LCR)$ models such that
\begin{equation} \label{constant-curvature conditions 0}
    f(0)=0,\myskip\myskip f'(0)=0,
\end{equation}
with the first condition being necessary for the model to harbour the vanishing-curvature solutions of GR. Analogously, \textit{every} metric with constant Ricci scalar $\LCR_0$ is a vacuum solution of \textit{any} $f(\LCR)$ model satisfying
\begin{equation} \label{constant-curvature conditions R0}
    f(\LCR_0)=0,\myskip\myskip f'(\LCR_0)=0.
\end{equation}
$f(\LCR)$ models fulfilling conditions \eqref{constant-curvature conditions 0} or \eqref{constant-curvature conditions R0} shall thus be the object of study of the present work. As explained in the introduction, we will generically refer to these special choices of function $f$ as \textit{$\LCR_0$-degenerate models}. In addition, we shall further distinguish between \textit{$(\LCR_0=0)$-degenerate models}, which satisfy conditions \eqref{constant-curvature conditions 0}, and \textit{$(\LCR_0\neq 0)$-degenerate models}, which comply with \eqref{constant-curvature conditions R0} instead. Finally, constant-curvature solutions exclusive to $\LCR_0$-degenerate $f(\LCR)$ models will be hereafter referred to as \textit{$\LCR_0$-degenerate} (in analogy with the models themselves), or as \textit{$f(\LCR)$-exclusive}, or even as \textit{novel} solutions.

The reader should note that it is straightforward to find non-trivial $\LCR_0$-degenerate $f(\LCR)$ models which appear to be, at least \textit{a priori}, physically well-motivated. The most paradigmatic examples would be the so-called `power-of-GR' models, $f(\LCR)\propto \LCR^{1+\delta}$, which fulfil conditions \eqref{constant-curvature conditions 0} provided that $\delta>0$. These models have interesting applications in cosmology and might be compatible with Solar System experiments, depending on the value of $\delta$ \cite{Clifton:2005aj,Bajardi:2022ocw}. Another simple instance of $(\LCR_0\neq 0)$-degenerate model would be $f(\LCR)=\LCR-\LCR_0/2-\LCR^2/(2\LCR_0)$, which satisfies \eqref{constant-curvature conditions R0} while behaving as $\LCR-2\Lambda+\bigo(\LCR^2)$ for $\LCR\ll\LCR_0$, being $\LCR_0$ the only dimensional parameter. The fact that there exist $\LCR_0$-degenerate models which reduce to GR$+\Lambda$ in the appropriate limit is a remarkable result, given that GR with (or without) a cosmological constant is not $\LCR_0$-degenerate by itself (for any $\LCR_0$).

Before closing this section, a brief comment on the predictability of $\LCR_0$-degenerate $f(\LCR)$ models is pertinent. As their name suggests, it is not clear whether $f(\LCR)$ models complying with either \eqref{constant-curvature conditions 0} or \eqref{constant-curvature conditions R0} have full predictive power. In any such $\LCR_0$-degenerate model, there is a set of initial conditions (namely, those requiring the Ricci scalar to be $\LCR_0$) whose evolution is not dictated by the vacuum equations of motion; recall equations \eqref{f(R) EOM} hold trivially for (the infinite number of) metrics with $\LCR=\LCR_0$. Moreover, there are indications that some metrics with the privileged Ricci scalar $\LCR_0$ can almost always be smoothly glued to each other \cite{Casado-Turrion:2022xkl}, thus suggesting that $\LCR_0$-degenerate models might be unable to discern between its (infinitely many) constant-curvature solutions.

\section{Strong-coupling pathologies in $\LCR_0$-degenerate $f(\LCR)$ models} \label{Section: Pathologies of the models}

Innocent as they might seem at first sight, the special class of $\LCR_0$-degenerate $f(\LCR)$ models---i.e.~those fulfilling either conditions \eqref{constant-curvature conditions 0} or \eqref{constant-curvature conditions R0}---can be shown to be inherently pathological, as stated in the introduction. In the following, we shall concentrate in a physically-relevant shortcoming of said models, namely, an apparent strong-coupling instability---i.e.~the non-propagation of all expected degrees of freedom at linear level around a flat background.

The linearised spectrum of a given gravity theory comprises all the independent fields which propagate on top of a suitable background when the equations of motion are expanded up to linear order in perturbations. The linearised spectrum around flat Minkowski spacetime thus coincides with the possible gravitational-wave polarisation modes which can be observationally detected, since the weak-field approximation is appropriate near current experimental settings.

It is well known that, generically, the gravitational wave spectrum of metric $f(\LCR)$ models consists of a massless and traceless graviton akin to that of GR (with two polarisation modes, the so-called `$+$' and `$\times$' polarisations) plus an additional longitudinal (i.e.~massive) scalar degree of freedom \cite{Capozziello:2008fn,Bouhmadi-Lopez:2012piq}, in consonance with the fact that $f(\LCR)$ theories of gravity are dynamically equivalent to a scalar-tensor theory \cite{Sotiriou:2008rp,DeFelice:2010aj}.\footnote{Some studies \cite{Alves:2009eg,Alves:2010ms,RizwanaKausar:2016zgi} claimed that the linearised spectrum of $f(\LCR)$ contained a second scalar polarisation mode, dubbed  \textit{breathing mode}, in disagreement with previous results. The controversy was finally settled against the existence of such a breathing mode resorting to the Hamiltonian formalism \cite{Liang:2017ahj} and gauge-invariant methods \cite{Moretti:2019yhs}.}

The fact that \textit{most} $f(\LCR)$ models propagate a massless and traceless graviton renders them compatible with gravitational wave observations \cite{Ezquiaga:2017ekz} (notice that no current gravitational-wave detectors are sensible to non-tensorial modes, including scalar modes, which remain unobserved). However, it is important to remark that, in general, previous analyses of gravitational waves in $f(\LCR)$ gravity made no assumptions on function $f$ itself (apart from analyticity at $\LCR=0$, which is necessary to perform the linearisation of the field equations, as we shall see later). For these reasons, these investigations failed to recognise that \textit{not all} $f(\LCR)$ models propagate the expected linearised degrees of freedom (graviton + scalar) around a Minkowski background. Indeed, we have found that $(\LCR_0=0)$-degenerate models feature such evanescence of the expected degrees of freedom, signalling the presence of a previously undiscovered strong-coupling instability in these models.\footnote{A background is said to be \textit{strongly-coupled} whenever at least one of the expected perturbative degrees of freedom fails to propagate atop said background. In other words, the kinetic term(s) of the evanescent field(s) vanish when evaluated in the strongly-coupled background; equivalently, interaction terms blow up upon canonicalisation of the equations of motion, hence the name \textit{strong-coupling}. Comprehensive accounts of the generalities of strong-coupling phenomena may be found in works investigating the appearance of such instabilities in physical theories. We refer the interested reader to references such as \cite{BeltranJimenez:2020lee}, for instance.} In particular, we have been able to establish the following two Results:

\begin{result} \label{Result: Non-propagation of gravity}
     At linear level in perturbations, $(\LCR_0=0)$-degenerate $f(\LCR)$ models---i.e.~those complying with conditions \eqref{constant-curvature conditions 0}---propagate, at most, one single massless scalar mode atop a flat background. In other words, these models do not contain the expected spin-2 graviton in their linearised spectrum, and thus Minkowski spacetime is strongly-coupled in $(\LCR_0=0)$-degenerate $f(\LCR)$ models.
\end{result}

\begin{result} \label{Result: Non-propagation of the scalaron}
    Around a Minkowski background, the linearised spectrum of $(\LCR_0=0)$-degenerate $f(\LCR)$ models satisfying $f''(0)=0$ does not contain any dynamical degrees of freedom whatsoever.
\end{result}

As mentioned before, Result \ref{Result: Non-propagation of gravity} puts into question the physical viability of $(\LCR_0=0)$-degenerate $f(\LCR)$ models, since the two polarisation modes corresponding to a massless and traceless spin-2 graviton have been detected in all gravitational-wave experiments carried out by the LIGO and VIRGO collaborations since 2015 \cite{LIGOScientific:2016aoc,LIGOScientific:2016lio}. We must also stress at this point that it is not very difficult to find $(\LCR_0=0)$-degenerate $f(\LCR)$ models that comply with the hypotheses of Result \ref{Result: Non-propagation of the scalaron}; for instance, all `power-of-GR' models $f(\LCR)\propto\LCR^{1+\delta}$ with $\delta>1$ (we shall also assume that $\delta$ is a natural number for the series expansion around $\LCR=0$ to exist).

In order to prove the assertions in Results \ref{Result: Non-propagation of gravity} and \ref{Result: Non-propagation of the scalaron}, we will proceed as follows. First, we will review the linearisation of the $f(\LCR)$ field equations \eqref{f(R) EOM} for any choice of function $f$. After that, we will particularise the results to the special case of $(\LCR_0=0)$-degenerate models---i.e.~we will make use of conditions \eqref{constant-curvature conditions 0}---to demonstrate the existence aforementioned apparent strong-coupling instabilities.

To perform the linear expansion of the $f(\LCR)$ equations of motion \eqref{f(R) EOM} around Minkowski spacetime, one starts by choosing a suitable coordinate system in which the metric $g_{\mu\nu}$ can be decomposed as
\begin{equation}
    g_{\mu\nu}=\eta_{\mu\nu}+h_{\mu\nu},
\end{equation}
where $\eta_{\mu\nu}$ is the Minkowski background and $h_{\mu\nu}$ is the metric perturbation, i.e.~$|h_{\mu\nu}|\ll 1$ in this special coordinate system. As widely known, the following expressions are true at first order in $h_{\mu\nu}$: 
\begin{eqnarray}
    g^{\mu\nu}&=&\eta^{\mu\nu}-h^{\mu\nu}+\bigo(h^2), \\
    \LCR_{\mu\nu}&=&\LCR_{\mu\nu}^{(1)}+\bigo(h^2), \\
    \LCR&=&\LCR^{(1)}+\bigo(h^2),
\end{eqnarray}
where
\begin{eqnarray}
    \LCR_{\mu\nu}^{(1)}&\equiv&\dfrac{1}{2}\left[\LCbox h_{\mu\nu}+\partial_\mu\partial_\nu h-2\partial_\lambda\partial_{(\mu}h_{\hphantom{\lambda}\nu)}^{\lambda}\right], \label{linearised array ini} \\
    \LCR^{(1)}&\equiv&\eta^{\mu\nu}R_{\mu\nu}^{(1)}=\LCbox h-\partial_\mu\partial_\nu h^{\mu\nu}, \label{R(1)} \\
    h^{\mu\nu}&\equiv&\eta^{\mu\rho}\eta^{\nu\sigma}h_{\rho\sigma}\,,\myskip
    h^\mu_{\hphantom{\lambda}\nu}\equiv\eta^{\mu\lambda}h_{\lambda\nu}\,,\myskip
    h\equiv\eta^{\mu\nu}h_{\mu\nu}.\myskip\myskip \label{linearised array fin}
\end{eqnarray}
In expressions \eqref{linearised array ini}--\eqref{linearised array fin} and hereafter, $\LCbox$ shall denote the Minkowski-space d'Alembertian, i.e.~$\LCbox=\eta^{\mu\nu}\partial_\mu\partial_\nu$.

The existence of a linearised regime of $f(\LCR)$ theories requires one additional assumption to be made, namely that $f$ must be analytic at $\LCR=0$, and therefore series-expandable up to to linear order in the Ricci-scalar perturbation $\LCR^{(1)}$, i.e.~up to linear order in metric perturbations. This is to guarantee that the resulting linearised equations of motion remain first-order in $h_{\mu\nu}$, something which is impossible whenever $f$ and its derivatives cannot be linearised in the first place.

For a generic $f(\LCR)$ theory of gravity which is analytic around $\LCR=0$, taking into account all the previous considerations results in the following set of linearised vacuum equations of motion:
\begin{equation} \label{general f(R) linearised EOM}
    f'(0)\,\LCEin_{\mu\nu}^{(1)}+f''(0)\,(\partial_\mu\partial_\nu-\eta_{\mu\nu}\LCbox)\LCR^{(1)}+\bigo(h^2)=0,
\end{equation}
where we have defined the Einstein-like tensor
\begin{equation} \label{Einstein-like tensor}
    \LCEin_{\mu\nu}^{(1)}\equiv\LCR_{\mu\nu}^{(1)}-\dfrac{1}{2}\eta_{\mu\nu}\LCR^{(1)}.
\end{equation}
Taking the trace of \eqref{general f(R) linearised EOM}, one finds
\begin{equation} \label{R(1) Klein-Gordon}
    3f''(0)\,\LCbox\LCR^{(1)}+f'(0)\LCR^{(1)}+\bigo(h^2)=0.
\end{equation}
which is a non-canonical Klein-Gordon equation for $\LCR^{(1)}$ provided that $f''(0)\neq 0$. Direct inspection of this equation clearly reveals that the kinetic term for $\LCR^{(1)}$ vanishes if $f''(0)=0$. Therefore, expression \eqref{R(1) Klein-Gordon} alone suffices to conclude that, only in cases where $f''(0)\neq 0$, the Ricci-scalar perturbation $R^{(1)}$ behaves as an independent, propagating scalar degree of freedom  at linearised level. One might then divide both sides of equation \eqref{R(1) Klein-Gordon} by $f''(0)$ so as to canonicalise the kinetic term, yielding
\begin{equation} \label{R(1) Klein-Gordon canonical}
    \LCbox\LCR^{(1)}+\dfrac{f'(0)}{3f''(0)}\LCR^{(1)}+\bigo(h^2)=0.
\end{equation}
Thus, for $f''(0)\neq 0$, the propagating scalar degree of freedom $\LCR^{(1)}$ has an effective mass $m_\mathrm{eff}$ given by
\begin{equation} \label{scalaron mass}
    m_\mathrm{eff}^2=\dfrac{f'(0)}{3f''(0)}.
\end{equation}

Turning back to the full linearised equations of motion \eqref{general f(R) linearised EOM}, we are now in a position that will allow us to understand intuitively why the spin-2 sector of the theory does not propagate atop the Minkowski background in $(\LCR_0=0)$-degenerate $f(\LCR)$ models. Aside from the higher-order terms, equation \eqref{general f(R) linearised EOM} contains (i) the term proportional to $f'(0)$ and the Einstein-like tensor $\LCEin^{(1)}_{\mu\nu}$ given by \eqref{Einstein-like tensor}, which encapsulates all terms depending on $h_{\mu\nu}$ and its derivatives; and (ii) the term containing derivatives of the scalar mode $R^{(1)}$, which is proportional to $f''(0)$ and does \textit{not} depend on $h_{\mu\nu}$ nor its derivatives. From this, it is clear that:
\begin{itemize}
    \item The first term (i) vanishes whenever function $f$ is such $f'(0)=0$, which is precisely one of the defining conditions of $(\LCR_0=0)$-generate $f(\LCR)$ models, cf.~\eqref{constant-curvature conditions 0}. Given that $\LCEin^{(1)}_{\mu\nu}$ contains all the derivatives of $h_{\mu\nu}$ appearing in the equations of motion, the absence of this term entails that the spin-2 mode does not propagate, as stated in Result \ref{Result: Non-propagation of gravity}. The strong-coupling problem becomes evident once one notices that the interaction terms---i.e.~the $\bigo(h^2)$ terms---blow up when one divides equation \eqref{general f(R) linearised EOM} by $f'(0)$ (in order to canonicalise the graviton kinetic terms) and then takes the limit $f'(0)\rightarrow 0$.
    \item The second term (ii) will not be present either whenever $f''(0)=0$, as discussed above. Thus, for $(\LCR_0=0)$-generate $f(\LCR)$ models such that $f''(0)=0$, equations \eqref{general f(R) linearised EOM} and \eqref{R(1) Klein-Gordon} contain no kinetic terms at all, only the $\bigo(h^2)$ interaction terms survive, and thus those theories do not possess a linearised spectrum, as asserted in Result \ref{Result: Non-propagation of the scalaron}.
\end{itemize}

However, there is a more insightful (and more mathematically explicit) way of understanding why the spin-2 degree of freedom fully decouples when $f'(0)=0$, i.e.~for $(\LCR_0=0)$-degenerate $f(\LCR)$ models. The argument goes as follows. As in GR, in $f(\LCR)$ gravity it is possible \cite{Capozziello:2008fn} to define a new symmetric rank-two tensor field, $\bar{h}_{\mu\nu}$, such that equations \eqref{general f(R) linearised EOM} reduce to the wave equation
\begin{equation} \label{GW equation}
    \LCbox\bar{h}_{\mu\nu}+\bigo(h^2)=0
\end{equation}
in the de Donder gauge, i.e.~after setting
\begin{equation}
    \partial_\mu\bar{h}^{\mu\nu}=0
\end{equation}
using some of the available gauge freedom in the theory. The remaining gauge freedom is then employed to impose the transverse-traceless (TT) condition. In a generic $f(R)$ theory of gravity, after expanding \eqref{general f(R) linearised EOM} in the de Donder gauge and comparing the result with \eqref{GW equation}, one finds that $\bar{h}_{\mu\nu}$ is given by\footnote{Our expression \eqref{hbar} for $\bar{h}_{\mu\nu}$ is slightly different from the one commonly found in the literature \cite{Capozziello:2008fn,RizwanaKausar:2016zgi},
\begin{equation*}
    \bar{h}_{\mu\nu}=\bar{h}_{\mu\nu}^\mathrm{GR}-\dfrac{f''(0)}{f'(0)}R^{(1)}\eta_{\mu\nu},
\end{equation*}
which, as the reader may immediately notice, is not valid if $f'(0)=0$, i.e.~precisely in the case we are interested in.
}
\begin{equation} \label{hbar}
    \bar{h}_{\mu\nu}=f'(0)\,\bar{h}_{\mu\nu}^\mathrm{GR}-f''(0)\,R^{(1)}\,\eta_{\mu\nu},
\end{equation}
where
\begin{equation}
    \bar{h}_{\mu\nu}^\mathrm{GR}\equiv h_{\mu\nu}-\dfrac{h}{2}\eta_{\mu\nu}
\end{equation}
is the usual spin-2 degree of freedom of GR. In consequence, equations \eqref{GW equation} and \eqref{hbar} evince that what propagates at the speed of light (in vacuum) in a generic $f(\LCR)$ gravity model is a mixture of the GR spin-2 graviton and the extra scalar mode. As per equation \eqref{hbar}, such propagating mixture reduces to its scalar component provided that $f'(0)=0$. In such situation, $R^{(1)}$ becomes effectively massless, due to \eqref{scalaron mass}, and the Klein-Gordon equation \eqref{R(1) Klein-Gordon} becomes equivalent to wave equation \eqref{GW equation}. As a result, only the massless scalar degree of freedom propagates in $(\LCR_0=0)$-degenerate $f(\LCR)$ models such that $f''(0)\neq 0$, as previously stated in Result \ref{Result: Non-propagation of gravity}. Again, one clearly sees that no propagating degree of freedom survives the limit $f''(0)\rightarrow 0$, in agreement with the strong-coupling instability described in Result \ref{Result: Non-propagation of the scalaron}.

\section{Stability of the novel constant-curvature solutions} \label{Section: Stability of solutions}

Even though $\LCR_0$-degenerate $f(\LCR)$ models possess an infinite number of solutions having constant scalar curvature $\LCR=\LCR_0$, it is actually possible to study the stability of all such solutions at once within a given model, without needing to perform a case-by-case analysis. In order to do so, we will resort to the Einstein-frame (i.e.~scalar-tensor) representation of $f(\LCR)$ gravities, which is ideally suited to study stability against small perturbations about a given constant value of $\LCR$.

As previously stated, it is well-known that metric $f(\LCR)$ theories can be regarded as equivalent to a scalar-tensor gravitational theory. More precisely, in the so-called Einstein frame \cite{Sotiriou:2008rp,DeFelice:2010aj},
\begin{equation} \label{Einstein-frame metric}
    \bar{g}_{\mu\nu}=f'(\LCR)\,g_{\mu\nu},
\end{equation}
the action \eqref{f(R) action} of metric $f(\LCR)$ gravity transforms into that of GR plus a dynamical gravitational scalar field $\phi$, with the latter being given by
\begin{equation} \label{scalaron}
    \phi(\LCR)=\sqrt{\dfrac{3}{2\kappa}}\ln f'(\LCR).
\end{equation}
This scalar field, also known as the \textit{scalaron}, is subject to the $f(\LCR)$-model-dependent potential
\begin{equation} \label{scalaron potential}
    V(\LCR)=\dfrac{f'(\LCR) \LCR-f(\LCR)}{2\kappa f'^2(\LCR)}.
\end{equation}
Notice that the Ricci scalar $\LCR$ appearing in the previous expressions is that of the so-called Jordan-frame metric, i.e.~the original, physical metric $g_{\mu\nu}$. As mentioned before, the scalaron \eqref{scalaron} is related to the scalar polarisation mode found in the gravitational-wave spectrum of the theory.

When working in the Einstein frame, the stability of a Jordan-frame constant-curvature solution will depend on whether $\LCR=\LCR_0$ is a minimum of the scalaron potential, and on whether such minimum is either global or local (in the latter case, the solution will only be metastable). Nonetheless, we must stress that some subtleties arise when using the Einstein frame in $\LCR_0$-degenerate models. The ones relevant to our work will be comprehensively discussed in Appendix \ref{Appendix: Caveats}. Also, it is straightforward to notice in \eqref{scalaron potential} that, if the $f(\LCR)$ model is $\LCR_0$-degenerate, a naive evaluation of $V(R)$ at $R=R_0$ leads to a $0/0$ indetermination. The reason is that, in such a case, both the numerator and the denominator in equation \eqref{scalaron potential} become zero when $R\rightarrow\LCR_0$, owing to conditions \eqref{constant-curvature conditions R0}. Therefore, the limit must be evaluated carefully.

There are two possible ways of computing limits which naively evaluate to indeterminations of the $0/0$ kind: (i) performing series expansions or (ii) applying L'H\^{o}pital's rule. The only difference between the aforementioned methods is their range of applicability: Taylor series require analyticity around the expansion point, while L'H\^{o}pital's rule only requires differentiability of the numerator and denominator. In our assessment of the stability of constant-curvature solutions in $\LCR_0$-degenerate models, we have made use of both methods, obtaining exactly the same outcomes, as shown in Results \ref{Result: R0 instability} and \ref{Result: R0=0 metastability} (recall that analyticity requires differentiability, and thus the results obtained L'H\^{o}pital's rule imply those obtained using series expansions). However, for the sake of clarity, and to avoid cluttering up this communication with long formulae, we shall only present the derivation using Taylor series, which produces shorter expressions at the expense of a more limited scope. However, we insist that the final results apply to non-analytic but differentiable $f$s as well.  It is worth noting that, regardless of method employed in the stability analysis, one is forced to assume that $f''(\LCR_0)\neq 0$.\footnote{Condition $f''(\LCR)\neq 0$ ensures that the correspondence between the Jordan and Einstein frames is well-posed \cite{Sotiriou:2008rp}. Moreover, the extrema of $V(\LCR)$ and $V(\phi)$ coincide if and only if $f''(\LCR)\neq 0$, as discussed in Appendix \ref{Appendix: Caveats}.}

Having explained why we have chosen to present just the computations using the series-expansion method, let us proceed with the stability analysis. As stated above, apart from demanding conditions \eqref{constant-curvature conditions R0} to hold (so that the model is $\LCR_0$-degenerate and harbours any solution with constant scalar curvature $\LCR_0$), we shall only make two additional assumptions, in particular, that function $f$ is analytic around $\LCR=\LCR_0$ (so that it can be Taylor-expanded around $\LCR_0$, as explained before), and also that $f''(R_0)\neq 0$. As a result, it is possible to expand both the numerator and the denominator of the scalaron potential \eqref{scalaron potential} around $\LCR=\LCR_0$. Indeed, close to $\LCR_0$, the denominator of $V(\LCR)$ behaves as
\begin{equation}
    \begin{split}
        f'^{-2}(\LCR) & \underset
    {\LCR=\LCR_0}{\sim}f''^{-2}(\LCR_0)(\LCR-\LCR_0)^{-2} \\
        & +\dfrac{f'''(\LCR_0)}{f''^3(\LCR_0)}(\LCR-\LCR_0)^{-1} \\
        & +\bigo[(\LCR-\LCR_0)^0].
    \end{split}
\end{equation}
whereas the numerator can be expanded as
\begin{equation}
    \begin{split}
        f'(\LCR) \LCR & - f(\LCR)\underset
        {\LCR=\LCR_0}{\sim} \LCR_0 f''(\LCR_0) (\LCR-\LCR_0) \\
        & +\dfrac{f''(\LCR_0)+\LCR_0 f'''(\LCR_0)}{2}(\LCR-\LCR_0)^2 \\
        & +\bigo[(\LCR-\LCR_0)^3].
    \end{split}
\end{equation}
As a result, we have that
\begin{multline} \label{potential expansion}
    2\kappa V(\LCR)\underset
    {\LCR=\LCR_0}{\sim}\dfrac{\LCR_0}{f''(\LCR_0)}(\LCR-\LCR_0)^{-1} \\
    +\dfrac{f''(\LCR_0)-\LCR_0 f'''(\LCR_0)}{2f''^2(\LCR_0)}+\bigo(\LCR-\LCR_0).
\end{multline}
We can now clearly infer from this last expression that the limit of $V(\LCR)$ as $\LCR\rightarrow\LCR_0$ does not exist unless $\LCR_0=0$. Certainly, should $\LCR_0$ be different from zero, series expansion \eqref{potential expansion} would be dominated by the order $(\LCR-\LCR_0)^{-1}$ term, which is a hyperbola tending to either positive or negative infinity depending whether $\LCR_0$ is approached from the left or the right. More precisely,
\begin{equation} \label{V limit at R=R0}
    \lim_{\substack{\LCR\rightarrow\LCR_0^\pm \\ \LCR_0\neq 0}}V(\LCR)=\sign\left[\dfrac{\LCR_0}{f''(\LCR_0)}\right]\times(\pm\infty).
\end{equation}
As mentioned earlier on this section, even though the previous limit has been computed using Taylor expansions (and thus under the assumption that $f$ is analytic at $\LCR=\LCR_0$), the equivalent computation using L'H\^{o}pital's rule yields exactly the same result \eqref{V limit at R=R0} for choices of $f$ which might not be analytic. We can therefore establish the following general result:

\begin{result} \label{Result: R0 instability}
Consider an $(\LCR_0\neq 0)$-degenerate $f(\LCR)$ model---i.e.~one fulfilling conditions \eqref{constant-curvature conditions R0}---such that $f''(\LCR_0)\neq 0$. Then its infinitely many solutions with constant curvature $\LCR=\LCR_0$ are generically unstable.
\end{result}

On the contrary, if $\LCR_0=0$, series expansion \eqref{potential expansion} yields
\begin{equation} \label{lim V R0 = 0}
    \lim_{\LCR\rightarrow 0}V(\LCR)=\dfrac{1}{4\kappa f''(0)},
\end{equation}
i.e.~the potential \eqref{scalaron potential} is analytic and perfectly well-defined at $\LCR=0$. Consequently, whenever $\LCR=0$ is a global minimum of the potential, zero-scalar-curvature solutions will be stable. If $\LCR=0$ is a local but non-global minimum, then the solutions will be just metastable (i.e.~stable only under small-enough perturbations). In either case, the additional constraints one must impose on $V(\LCR)$ are the usual minimum conditions:
\begin{eqnarray}
    V'(0)&=&\lim_{\LCR\rightarrow 0} V'(R)=0, \label{Minimum Condition 1} \\
    V''(0)&=&\lim_{\LCR\rightarrow 0} V''(R)>0\,\text{ but finite}. \label{Minimum Condition 2}
\end{eqnarray}
Again, should the limits of $V'(\LCR)$ and $V''(\LCR)$ as $\LCR$ tends to zero be taken directly, indeterminations of the $0/0$-type would emerge in each case. Consequently, one has to proceed exactly as we have done above when evaluating the limit of the potential itself. As before, we shall only assume that $f$ is analytic around $\LCR=\LCR_0=0$, and that $f''(0)\neq 0$.

Taking this into account, the limit of $V'(\LCR)$ as $\LCR\rightarrow 0$ may be again computed by expanding its numerator and denominator around $\LCR=0$, yielding
\begin{equation} \label{lim V' R0 = 0}
    \lim_{\LCR\rightarrow 0}V'(\LCR)=-\dfrac{f'''(0)}{12\kappa f''^2(0)}. \\
\end{equation}
As per local minimum condition \eqref{Minimum Condition 1}, metastability of the solutions with $\LCR=0$ then requires
\begin{equation}
    f'''(0)=0.
\end{equation}
Similarly, one finds that the limit of $V''(\LCR)$ as $\LCR\rightarrow 0$ is also well-defined, being
\begin{equation}
    \lim_{\LCR\rightarrow 0}V''(\LCR)=-\dfrac{f''''(0)}{24\kappa f''^2(0)}.
\end{equation}
As a result, local minimum condition \eqref{Minimum Condition 2} implies that an $(\LCR_0=0)$-degenerate $f(\LCR)$ model must be such that
\begin{equation} \label{last minimum condition on f}
    f''''(0)<0
\end{equation}
for its infinitely many vanishing-scalar-curvature solutions to be at least metastable. Once more, as in the $\LCR_0\neq 0$ case, results \eqref{V limit at R=R0} and \eqref{lim V' R0 = 0}--\eqref{last minimum condition on f} are also found using L'H\^{o}pital's rule, and therefore hold for functions $f$ which need not be analytic. In consequence, we can formulate the following generic result:

\begin{result} \label{Result: R0=0 metastability}
    All vanishing-scalar-curvature spacetimes are at least metastable vacuum solutions of $(\LCR_0=0)$-degenerate $f(\LCR)$ models---i.e.~those satisfying conditions \eqref{constant-curvature conditions 0}---provided that function $f$ is such that (i) $f''(0)\neq 0$, (ii) $f'''(0)=0$, and (iii) $f''''(0)<0$.
\end{result}

Analytic $f(\LCR)$ models satisfying all the assumptions of Result \ref{Result: R0=0 metastability} would admit the following Taylor expansion around $\LCR=0$:
\begin{equation}
    f(\LCR)=\alpha\LCR^2-\beta\LCR^4+\bigo(\LCR^5),
\end{equation}
where $\alpha\neq 0$ and $\beta>0$ are real constants. The simplest such models are, evidently, the polynomic ones with
\begin{equation} \label{stable R = 0 polynomial f(R)}
    f(\LCR)=\alpha\LCR^2-\beta\LCR^4.
\end{equation}
It is not difficult to show that the potential associated to \eqref{stable R = 0 polynomial f(R)} has a global minimum at $\LCR=0$ provided that $\alpha<0$; in such case, however, the Dolgov-Kawasaki stability condition $f''(\LCR)>0$ \cite{Dolgov:2003px,Seifert:2007fr} is violated. On the other hand, if $\alpha>0$, the Dolgov-Kawasaki condition holds, but then $\LCR=0$ is a only a local minimum of the potential. The only other extrema of $V(\LCR)$ are two maxima at $\LCR=\pm\sqrt{\alpha/6\beta}$. Thus, for $|\LCR|>\sqrt{\alpha/6\beta}$ the potential is monotonically decreasing, i.e.~$V(\LCR)\rightarrow-\infty$ as $\LCR\rightarrow\pm\infty$. In other words, the model has a potential not bounded below and lacks a ground state. Thus, the metastable solutions with $R=0$ could be driven to infinite scalar curvature, should the perturbations applied around such solutions be large enough.

\section{Some paradigmatic $f(R)$-exclusive constant-curvature solutions} \label{introducing the solutions}

As we saw before in Section \ref{section:constant-curvature solutions}, \textit{any} spacetime with constant curvature $\LCR_0$ is a solution of \textit{any} $\LCR_0$-degenerate $f(\LCR)$ model. Therefore, the problem of obtaining new constant-curvature solutions for such $\LCR_0$-degenerate models reduces to solving the differential equation $\LCR=\LCR_0$. This can be accomplished by postulating several simple ans\"{a}tze for the metric.

As a first approximation to the problem, one may require the constant-curvature solution to be static and spherically symmetric. In such case, it is always possible to choose `areal-radius' coordinates $(t,r,\theta,\varphi)$ such that the most generic static and spherically symmetric line element can be written as
\begin{equation} 
\label{static spherically symmetric}
    \dif s^2=A(r)\,\dif t^2-B(r)\,\dif r^2-r^2\,\dif\Omega^2,
\end{equation}
where $A$ and $B$ are the only independent metric functions. Line element \eqref{static spherically symmetric} has Ricci scalar
\begin{eqnarray} \label{static spherically symmetric Ricci scalar}
    \LCR=&-&\dfrac{A''}{AB}+\dfrac{A'}{2AB}\left(\dfrac{A'}{A}+\dfrac{B'}{B}\right) \nonumber \\
    &&-\,\dfrac{2}{r}\left(\dfrac{A'}{AB}-\dfrac{B'}{B^2}\right)+\dfrac{2}{r^2}\left(1-\dfrac{1}{B}\right).
\end{eqnarray}
Notice that \eqref{static spherically symmetric} is directly expressed in the so-called Abreu-Nielsen-Visser gauge,
\begin{eqnarray}
    \dif s^2 &=& \e^{2\Phi}\left(1-\dfrac{2GM_\mathrm{MSH}}{r}\right)\dif t^2 \nonumber \\
    &&-\,\left(1-\dfrac{2GM_\mathrm{MSH}}{r}\right)^{-1}\dif r^2-r^2\,\dif\Omega^2, \label{Abreu-Nielsen-Visser}
\end{eqnarray}
with Misner-Sharp-Hern\'{a}ndez (MSH) mass \cite{Misner:1964je,Hernandez:1966zia}
\begin{equation} \label{MSH mass}
    M_\mathrm{MSH}(r)=\dfrac{r}{2G}\left(1-\dfrac{1}{B(r)}\right)
\end{equation}
and anomalous redshift function
\begin{equation} \label{anomalous redshift}
    \Phi(r)=\dfrac{1}{2}\ln\left[A(r)B(r)\right].
\end{equation}
These two quantities will help us interpret the various solutions to be discussed in what follows.

Equation $\LCR=\LCR_0=\const$---with $\LCR$ given by expression \eqref{static spherically symmetric Ricci scalar}---has infinitely many solutions, i.e.~there is an infinite number of static, spherically symmetric spacetimes having constant Ricci scalar.\footnote{A remarkable observation is that, given any function $A(r)$, there is a closed-form expression for the precise function $B(r)$ which solves equation $\LCR=\LCR_0$, namely
\begin{equation*}
    B(r)=\dfrac{I(r)}{B_0+B_1(r)},
\end{equation*}
where $B_0$ is an integration constant,
\begin{equation*}
    B_1(r)\equiv\int_1^r\dfrac{\dif x}{x}\dfrac{4I(x)A(x)}{4A(x)+xA'(x)}\left(1-\dfrac{\LCR_0}{2}x^2\right),
\end{equation*}
and function $I(r)$ is defined as
\begin{equation*}
    I(r)\equiv\exp\left[\int_1^r\dfrac{\dif x}{x}\dfrac{4A(x)+4xA'(x)-\frac{x^2A'^2(x)}{A(x)}+2x^2A''(x)}{4A(x)+xA'(x)}\right].
\end{equation*}
} For this reason, we will focus on spherically symmetric spacetimes which have been previously reported in the literature \cite{Calza:2018ohl,Casado-Turrion:2022xkl}, or generalisations thereof. Once we have presented the paradigmatic solutions in Sub-Sections \ref{Section: Class 1}--\ref{Section: Class 3}, we will proceed to characterise them. That is to say, we would like to know which kind of objects such solutions represent, and whether they have physically meaningful properties. With that purpose, we will analyse the following five aspects:
\begin{itemize}
    \item Apparent and Killing horizons.
    \item Coordinate singularities.
    \item Curvature singularities.
    \item Geodesic completeness, i.e.~whether the curvature singularities can be reached in finite time by freely falling observers.
    \item Regions in which the metric has an unphysical signature (e.g.~regions in which there are two time coordinates or the metric becomes Euclidean), and whether these pathological regions can be reached in finite time by freely-falling observers.
\end{itemize}
The precise definitions of the properties listed above may be found on Appendix \ref{Appendix: Characterisation}, along with useful formulae that will be employed during our characterisation of the solutions. For instance, equation \eqref{Delta lambda} will allow us to determine whether a singularity or a region with unphysical metric signature is out of reach for causal observers.

Our findings concerning the points above are summarised in Table \ref{tab:solutions}. As may be deduced from these results, the paradigmatic solutions to be described in what follows exhibit a number of unusual characteristics which put their physical viability into question. As said before, given that the number of constant-curvature solutions a certain $\LCR_0$-degenerate model hosts is infinite, one expects the majority of those constant-curvature spacetimes to be pathological. This Section is thus intended to provide a limited but illustrative picture of the kind of issues one should expect to find when dealing with novel, $f(\LCR)$-exclusive constant-curvature solutions.

\begin{table*}
\caption{\label{tab:solutions}%
Overview of the constant-curvature solutions considered in this work and their pathological characteristics.
}
\begin{minipage}{\textwidth}
\begin{tabularx}{\textwidth}{ccccL}
\hline\hline
\multirow{2}{*}{Type}&
Parameters &
\multirow{2}{*}{Subclass}&
\multirow{2}{*}{Line element}&
\multicolumn{1}{c}{\multirow{2}{*}{Issues and oddities}} \\
& (dimensions) & & & \\
\colrule
\multirow{10}{*}{Class 1} & \multirow{10}{*}{$C\neq 0$ (L)} & \multirow{4}{*}{$C>0$} & \multirow{4}{*}{\eqref{Class 1.1}} & \tabitemize{\item Protected curvature singularity at the origin $r=0$ (i.e.~it cannot be reached by causal observers in finite proper time). \item Can be smoothly matched with an interior Minkowski spacetime at any given radius for every $C$ \cite{Casado-Turrion:2022xkl}.} \\ \cline{3-5}
& & \multirow{6}{*}{$C<0$} & \multirow{6}{*}{\eqref{Class 1.2}} & \tabitemize{\item Protected curvature singularity at the origin $r=0$. \item Accessible curvature singularity at $r=|C|$ (i.e.~causal observers can reach it in finite proper time). \item Surface $r=|C|$ also appears to be a Killing horizon, but it is not null. \item As their counterparts with $C<0$, they can be smoothly matched with an interior Minkowski spacetime at any given radius for every $C$ \cite{Casado-Turrion:2022xkl}.} \\ \hline
\multirow{9}{*}{Class 2} & \multirow{9}{*}{\shortstack[c]{$\LCR_0$ ($\mathrm{L}^{-2}$)\footnote{\label{footnote: table R0}Notice that, by changing the value of $\LCR_0$, one is actually changing the set of $f(\LCR)$ models in which this spacetime is a vacuum solution, cf.~conditions \eqref{constant-curvature conditions 0} and \eqref{constant-curvature conditions R0}.} \\ $D$ ($\mathrm{L}^0$) \\ $C\neq 0$ ($\mathrm{L}$)\footnote{As explained in Section \ref{Section: Class 2} in the bulk of the text, this constant can always be absorbed by a redefinition of the time coordinate. We have deliberately kept it in the metric for purely dimensional purposes (i.e.~so as to have a time coordinate with dimensions of length), and also to explicitly demonstrate that it represents the same physical quantity as similarly-named parameter $C$ in Class 1 solutions.}}} & \multirow{4}{*}{$\LCR_0>0$} & \multirow{4}{*}{\eqref{Class 2.1}} & \tabitemize{\item Infinite number of accessible curvature singularities. \item Infinite number of would-be (non-null) Killing horizons, all coincident with curvature singularities. \item Generically unstable, as per Result \ref{Result: R0 instability}.} \\ \cline{3-5}
& & \multirow{2}{*}{$\LCR_0<0$} & \multirow{2}{*}{\eqref{Class 2.2}} & \tabitemize{\item Protected curvature singularity at the origin. \item Generically unstable, as per Result \ref{Result: R0 instability}.} \\ \cline{3-5}
& & \multirow{3}{*}{$\LCR_0=0$\footnote{In this case, dimensionless constant $D$ can always be absorbed by a coordinate redefinition.}} & \multirow{3}{*}{\eqref{Class 2.0}} & \tabitemize{\item Protected curvature singularity at the origin. \item Describes a pair of disconnected parallel universes, one at each side of the central singularity.} \\ \hline
\multirow{35}{*}{Class 3} & \multirow{35}{*}{\shortstack[c]{$\LCR_0$ ($\mathrm{L}^{-2}$)\footnote{See footnote \ref{footnote: table R0}.} \\ $M$ ($\mathrm{L}^{-1}$)\footnote{$\LCR_0$ and $M$ cannot vanish at the same time because, in that case, Class 3 solutions would trivially reduce to Minkowski spacetime.}}} & \multirow{6}{*}{$M=0$, $\LCR_0>0$} & \multirow{6}{*}{\eqref{Class 3 M = 0 R0 > 0}} & \tabitemize{\item The metric signature becomes unphysical, i.e.~$(+,+,-,-)$, for radii $r$ larger than the critical value $r=\sqrt{6/\LCR_0}$. \item Causal observers can reach the region with unphysical metric signature in finite proper time. \item $r=\sqrt{6/\LCR_0}$ is also an apparent horizon (a null surface in which $g^{rr}=0$). \item Generically unstable, as per Result \ref{Result: R0 instability}.} \\ \cline{3-5}
& & \multirow{2}{*}{$M=0$, $\LCR_0<0$} & \multirow{2}{*}{\eqref{Class 3 M = 0 R0 < 0}} & \tabitemize{\item Describes a traversable wormhole made out of pure vacuum. \item However, it is generically unstable, as per Result \ref{Result: R0 instability}.} \\ \cline{3-5}
& & \multirow{6}{*}{$M>0$, $\LCR_0=0$} & \multirow{6}{*}{\eqref{Class 3 M > 0 R0 = 0}} & \tabitemize{\item The metric signature becomes unphysical for $r<2GM$. \item Causal observers can reach the region with unphysical metric signature in finite proper time. \item $r=2GM$ is also the location of an apparent horizon. \item There is a curvature singularity located at $r=0$ (i.e.~within the unphysical region).} \\ \cline{3-5}
& & $M<0$, $\LCR_0=0$ & \eqref{Class 3 M < 0 R0 = 0} & \tabitemize{\item Accessible curvature singularity at the origin $r=0$.} \\ \cline{3-5}
& & \multirow{2}{*}{$M<0$, $\LCR_0<0$} & \multirow{2}{*}{\eqref{Class 3 M < 0 R0 < 0}} & \tabitemize{\item Accessible curvature singularity at the origin $r=0$. \item Generically unstable, as per Result \ref{Result: R0 instability}.} \\ \cline{3-5}
& & \multirow{5}{*}{$M>0$, $\LCR_0<0$} & \multirow{5}{*}{\eqref{Class 3 M > 0 R0 < 0}} & \tabitemize{\item The metric signature becomes unphysical for radii $r$ smaller than some critical value $r_\mathrm{ah}$ given by expression \eqref{Apparent horizon Class 3 M > 0 R0 < 0}. \item There is a curvature singularity located at $r=0$ (i.e.~within the unphysical region). \item Generically unstable, as per Result \ref{Result: R0 instability}.} \\ \cline{3-5}
& & \multirow{4}{*}{$M<0$, $\LCR_0>0$} & \multirow{4}{*}{\eqref{Class 3 M < 0 R0 > 0}} & \tabitemize{\item The metric signature becomes unphysical for radii $r$ larger than some critical value $r_\mathrm{ah}$ given by expression \eqref{Apparent horizon Class 3 M < 0 R0 > 0}. \item There is a curvature singularity located at $r=0$. \item Generically unstable, as per Result \ref{Result: R0 instability}.} \\ \cline{3-5}
& & \multirow{10}{*}{$M>0$, $\LCR_0>0$} & \multirow{10}{*}{\eqref{Class 3 M > 0 R0 > 0}} & \tabitemize{\item If $3GM\sqrt{\LCR_0/2}\geq 1$, the metric signature is unphysical for all $r$. If $3GM\sqrt{\LCR_0/2}=1$, then $r=3GM=\sqrt{2/\LCR_0}$ would be a wormhole throat (were the metric signature physical). \item If $0<3GM\sqrt{\LCR_0/2}<1$, the metric signature is unphysical for $r<r_0$ and for $r>r_1$, where $r_0$ and $r_1$ are two critical values given by expressions \eqref{Apparent horizon Class 3 M > 0 R0 > 0 r_0} and \eqref{Apparent horizon Class 3 M > 0 R0 > 0 r_1}, respectively. Surfaces $r=r_0$ and $r=r_1$ correspond to apparent horizons. \item There is a curvature singularity located at $r=0$ (i.e.~within the region with unphysical metric signature). \item Generically unstable, as per Result \ref{Result: R0 instability}.} \\
\hline\hline
\end{tabularx}
\end{minipage}
\end{table*}

\subsection{Novel solutions of Class 1} \label{Section: Class 1}

The first set of $f(\LCR)$-exclusive constant-curvature solutions we would like to discuss (hereafter to be known as Class 1 solutions) shall be those metrics whose line element can be expressed as
\begin{equation} \label{Class 1.1}
    \dif s^2=\left(1+\dfrac{C}{r}\right)^2\dif t^2-\dif r^2-r^2\dif\Omega^2,
\end{equation}
where $C$ is a free parameter with dimensions of length, which might be either positive or negative. These spacetimes have vanishing Ricci scalar for any value of $C$, and thus solve the equations of motion of all $f(\LCR)$ models complying with conditions \eqref{constant-curvature conditions 0}. An interesting property of Class 1 spacetimes is that they can be smoothly matched to a Minkowski interior at any given spherical surface $r=r_*=\const$, as shown in \cite{Casado-Turrion:2022xkl}. Metric \eqref{Class 1.1} thus represents spacetime outside of such a static vacuole solution.

The most straightforward way of obtaining line element \eqref{Class 1.1} consists in setting $\LCR=0$ in equation \eqref{static spherically symmetric Ricci scalar} and solving for $A(r)$ using the simple ansatz $B(r)=1$. By doing so, one finds that the general solution is given by the (a priori) two-parameter family of metrics
\begin{equation} \label{Class 1 general solution}
    \dif s^2=\left(D+\dfrac{C}{r}\right)^2\dif t^2-\dif r^2-r^2\dif\Omega^2,
\end{equation}
where $C$ and $D$ are, again, real constants, with $D$ being dimensionless. However, it is not difficult to realise that parameter $D$ can always be removed from the metric. If $D\neq 0$, it is always possible to transform line element \eqref{Class 1 general solution} into the Class 1 form \eqref{Class 1.1} presented above by redefining $C/D\rightarrow C$ and $D^2\,\dif t^2\rightarrow\dif t^2$. Nonetheless, if $D=0$, metric \eqref{Class 1 general solution} becomes
\begin{equation} \label{Class 2.0}
    \dif s^2=\left(\dfrac{C}{r}\right)^2\dif t^2-\dif r^2-r^2\dif\Omega^2,
\end{equation}
Spacetimes of this form \eqref{Class 2.0} are a particular instance of one of the families of $f(\LCR)$-exclusive constant-curvature solutions originally discovered by Calz\`{a}, Rinaldi and Sebastiani in Reference \cite{Calza:2018ohl}.\footnote{More precisely, line element \eqref{Class 2.0} can be obtained by setting $b=2$, $z=-2$ and $c_0=0$ in equations (33)--(35) of \cite{Calza:2018ohl}.} We shall consider metrics of the form \eqref{Class 2.0} to be solutions of Class 2, given that they cannot be recovered from the standard Class 1 form \eqref{Class 1.1}, and for further reasons that will become apparent in subsection \ref{Section: Class 2}. Consequently, in the present subsection we shall focus on analysing only the properties of metrics of the form \eqref{Class 1.1}.

\subsubsection{General properties of Class 1 solutions}

It is clear from line element \eqref{Class 1.1} that Class 1 solutions are asymptotically flat, and that they also reduce to Minkowski spacetime in the limit $C\rightarrow 0$. Using \eqref{MSH mass} and \eqref{anomalous redshift}, one might immediately deduce that metrics of the form \eqref{Class 1.1} have vanishing MSH mass and anomalous redshift function
\begin{equation} \label{vacuole exterior redshift}
    \Phi(r)=\ln\left|1+\dfrac{C}{r}\right|.
\end{equation}
In consequence, Class 1 solutions may be regarded as metrics deviating from Minkowski spacetime only in a gravitationally-induced anomalous redshift. The interpretation of length scale $C$ appearing in \eqref{Class 1.1} is far less clear. As shown in \cite{Casado-Turrion:2022xkl}, the value of this parameter cannot be fixed by glueing \eqref{Class 1.1} to a Minkowski spacetime at a given spherical surface $r=r_*=\const$ Class 1 solutions are also strikingly similar to the extremal Reissner-Nordstr\"{o}m spacetime, sharing a common Newtonian limit $g_{tt}\simeq 1+2\phi_\mathrm{N}$ (where $\phi_\mathrm{N}$ is the Newtonian potential) upon identification of $C$ with (Newton's constant times) the mass or the charge of the black hole. In spite of this, Class 1 metrics are not sourced by any electromagnetic field. Instead, as mentioned before, the correction to the point-like-mass potential in the weak field limit of \eqref{Class 1.1} is entirely due to anomalous gravitational effects (recall this is a vacuum solution). Therefore, the interpretation of $C$ as an `extremal charge' seems not to be appropriate.

Expressions \eqref{Class 1.1} and \eqref{vacuole exterior redshift} reveal that the remaining properties of Class 1 solutions depend crucially on the sign of $C$. As such, we will study separately the sub-class of solutions with $C>0$ and the sub-class with $C<0$.

\subsubsection{Class 1 solutions with $C>0$}

Direct inspection of line element \eqref{Class 1.1} reveals that Class 1 solutions with $C>0$ only harbour one coordinate singularity at $r=0$, with the metric remaining regular for any other $r>0$. Moreover, when $C>0$ there are neither apparent or Killing horizons, nor any regions in which the signature of line element \eqref{Class 1.1} becomes unphysical.

Given that the Kretschmann scalar corresponding to \eqref{Class 1.1} is
\begin{equation}
    \LCKre=\dfrac{24 C^2}{r^6}\left(1+\dfrac{C}{r}\right)^{-2},
\end{equation}
one immediately realises that the coordinate singularity at the origin is an actual curvature singularity. Despite this, the singularity is inaccessible for any causal observer. Integral \eqref{Delta lambda} may be computed for Class 1 solutions with $C>0$ directly, yielding
\begin{equation}
    \Delta\lambda(r_\mathrm{ini}\rightarrow r_\mathrm{fin})=\left|r_\mathrm{fin}-r_\mathrm{ini}+C\ln\left(\dfrac{r_\mathrm{fin}}{r_\mathrm{ini}}\right)\right|.
\end{equation}
It is thus apparent that, along a radial null geodesic, the affine parameter separating any $r>0$ and the central singularity is infinite. Because photons take an infinite amount of affine parameter to reach the central singularity, no other particle can reach $r=0$ in finite proper time either. In consequence, for all practical purposes, the singularity at $r=0$ is inaccessible, and spacetime \eqref{Class 1.1} is geodesically complete for $C>0$.

The previous result can be easily understood in view of the form of the anomalous redshift function \eqref{vacuole exterior redshift} for Class 1 spacetimes with $C>0$. As we can clearly deduce from this expression, the anomalous redshift becomes infinite at $r=0$. In other words, modified-gravity effects induce an ever-increasing redshift function which \textit{protects} observers from the curvature singularity.

\subsubsection{Class 1 solutions with $C<0$}

The metric of Class 1 spacetimes with $C<0$, viz.
\begin{equation} \label{Class 1.2}
    \dif s^2=\left(1-\dfrac{|C|}{r}\right)^2\dif t^2-\dif r^2-r^2\dif\Omega^2,
\end{equation}
exhibits two coordinate singularities: the one at $r=0$ (which was already present in the $C>0$ sub-class), and a second one at $r=|C|$. Both prove to be curvature singularities, since the Kretschmann scalar becomes
\begin{equation}
    \LCKre=\dfrac{24C^2}{r^6}\left(1-\dfrac{|C|}{r}\right)^{-2}
\end{equation}
for $C<0$. As we shall see now, the singularity at $r=0$ remains inaccessible for radially infalling photons. However, the singularity at $r=|C|$ is causally connected to the rest of the spacetime.

On the one hand, for $r_\mathrm{ini}$, $r_\mathrm{fin}<|C|$, integral \eqref{Delta lambda} yields
\begin{equation}
    \Delta\lambda(r_\mathrm{ini}<|C|\rightarrow r_\mathrm{fin}<|C|)=\left|r_\mathrm{fin}-r_\mathrm{ini}-|C|\ln\left(\dfrac{r_\mathrm{fin}}{r_\mathrm{ini}}\right)\right|.
\end{equation}
One may readily substitute $r_\mathrm{fin}=0$ in this expression to find that the curvature singularity at $r=0$ cannot be reached in finite time by causal observers. In parallel with the $C>0$ case, the curvature singularity is protected by an $f(\LCR)$-induced infinite redshift.

On the other hand, if $r_\mathrm{ini}\neq |C|$, but $r_\mathrm{fin}=|C|$, then integral \eqref{Delta lambda} becomes
\begin{equation}
    \Delta\lambda(r_\mathrm{ini}\rightarrow |C|)=\left||C|-r_\mathrm{ini}-|C|\ln\left(\dfrac{|C|}{r_\mathrm{ini}}\right)\right|,
\end{equation}
which is finite for every $r_\mathrm{ini}\neq 0$. Therefore, a photon travelling along a null radial geodesic will reach the singularity at $r=|C|$ in finite time, should it not be `emitted' at the other singularity at the origin.

For completeness, we shall also compute $\Delta\lambda(r_\mathrm{ini}\rightarrow r_\mathrm{fin})$ for $r_\mathrm{ini}>|C|$ and $r_\mathrm{fin}\leq|C|$ (without loss of generality\footnote{As mentioned in Appendix \ref{Appendix: Characterisation}, the absolute value in \eqref{Delta lambda} entails that $\Delta\lambda(r_\mathrm{ini}\rightarrow r_\mathrm{fin})=\Delta\lambda(r_\mathrm{fin}\rightarrow r_\mathrm{ini})$, so it is always possible to exchange $r_\mathrm{ini}\leftrightarrow r_\mathrm{fin}$.}). To do so, we must first remark that
%\begin{equation}
%    \sqrt{A(r)B(r)}=\left|1-\dfrac{|C|}{r}\right|,
%\end{equation}
%which equals $(1-|C|/r)$ if $r>|C|$ and $-(1-|C|/r)$ if $r<|C|$. Thus,
the integral in \eqref{Delta lambda} breaks into two pieces in this case:
\begin{multline}
    \Delta\lambda(r_\mathrm{ini}\rightarrow r_\mathrm{fin}) \\
    =\left|\int_{r_\mathrm{ini}}^{|C|}\dif r\,\left(1-\dfrac{|C|}{r}\right)-\int_{|C|}^{r_\mathrm{fin}}\dif r\,\left(1-\dfrac{|C|}{r}\right)\right|.
\end{multline}
As a result,
\begin{equation}
    \Delta\lambda(r_\mathrm{ini}\rightarrow r_\mathrm{fin})=\left|r_\mathrm{ini}+r_\mathrm{fin}-2|C|\left[1+\ln\left(\dfrac{\sqrt{r_\mathrm{ini}\,r_\mathrm{fin}}}{|C|}\right)\right]\right|.
\end{equation}
This expression attests once more that, whatsoever value $r_\mathrm{ini}$ takes, $\Delta\lambda(r_\mathrm{ini}\rightarrow r_\mathrm{fin})$ remains finite unless $r_\mathrm{fin}=0$. Thus, the central singularity is protected, by the one at $r=|C|$, the latter being truly naked.

Another surprising feature of surface $r=|C|$ is that it appears to be a Killing horizon for $\xi=\partial/\partial t$, since
\begin{equation}
    g_{\mu\nu}\xi^\mu\xi^\nu=\left(1-\dfrac{|C|}{r}\right)^2
\end{equation}
vanishes at $r=|C|$. However, $r=|C|$ is not a null surface, because its normal vector $n_\mu=\partial_\mu r=\delta^r_{\hphantom{r}\mu}$ is everywhere timelike. Therefore, strictly speaking, it cannot be a Killing horizon.\footnote{Given that $r=|C|$ is also a curvature singularity, as seen before, it is technically not part of the spacetime.} Given the fact that $g_{\mu\nu}\xi^\mu\xi^\nu>0$ for any $r\neq|C|$, the would-be Killing horizon at $r=|C|$ would be degenerate, in analogy with the true Killing horizon of extremal Reissner-Nordstr\"{o}m black holes. A remarkable difference between Class 1 solutions with $C<0$ and extremal Reissner-Nordstr\"{o}m black holes is that the former do not possess any apparent horizons, while the Killing horizon of extremal Reissner-Nordstr\"{o}m black holes is also an apparent horizon.\footnote{Furthermore, the horizon of extremal Reissner-Nordstr\"{o}m black holes is regular, i.e.~free of curvature singularities.}

\subsection{Novel solutions of Class 2} \label{Section: Class 2}

Class 2 shall encompass three different kinds of constant-curvature solutions, all of which are characterised by their compliance with the simple ansatz $B(r)=1$, as those of Class 1. As a result, all Class 2 spacetimes will have vanishing MSH mass, and thus only differ from Minkowski spacetime in an $f(\LCR)$-induced anomalous redshift factor.

The first two sub-classes within Class 2 will contain, respectively, the solutions of equation $\LCR=\LCR_0$ with $B(r)=1$ and either positive (first subclass) or negative (second subclass) constant curvature $R_0$ ---recall that the Ricci scalar of static spacetimes is given by expression \eqref{static spherically symmetric Ricci scalar}---. The third subclass will be conformed by those spacetimes resulting from setting $\LCR_0\rightarrow 0$ in the aforementioned Class 2 solutions with $\LCR_0\neq 0$.

As done in the previous section, we will analyse each subclass within Class 2 separately.

\subsubsection{Class 2 solutions with $\LCR_0>0$}

Substituting the simple ansatz $B(r)=1$ in \eqref{static spherically symmetric Ricci scalar} and solving the equation $\LCR=\LCR_0>0$ for $A(r)$, one finds the following family of constant-curvature solutions of $(\LCR_0\neq 0)$-degenerate $f(\LCR)$ gravity models:
\begin{equation} \label{Class 2.1}
    \dif s^2=\left(\dfrac{C}{r}\right)^2\cos^2\left(\sqrt{\dfrac{\LCR_0}{2}}\,r+D\right)\dif t^2-\dif r^2-r^2\dif\Omega^2,
\end{equation}
where $C\neq 0$ and $D$ are integration constants, $C$ having dimensions of length, and $D$ being dimensionless. Notice that, even though $C$ can always be removed from the line element through the coordinate redefinition $t\rightarrow t/C$, we have kept it in \eqref{Class 2.1} for dimensional reasons (i.e.~the new, re-scaled time coordinate would be dimensionless, while the new $g_{tt}$ would have dimensions of length squared).

Class 2 spacetimes with $\LCR_0>0$ are highly pathological; despite their simple appearance, they are, perhaps, the least physically well-founded solutions we will consider in this work. For instance, it is immediate to infer from expression \eqref{Class 2.1} that these solutions have an infinite number of coordinate singularities. One of them is located at $r=0$, where $g_{tt}\rightarrow\infty$, while the remaining ones are located at the points in which the cosine vanishes, that is to say, at
\begin{equation} \label{r_n}
    r_n=\sqrt{\dfrac{2}{\LCR_0}}\left[\dfrac{\pi}{2}(2n+1)-D\right],
\end{equation}
with $n$ an integer.\footnote{We remark that the allowed values of $n$ depend on the value of $D$. In particular, the minimum possible $n$ should be such that $r_n>0$ for all permitted $n$s.} Direct computation of the Kretschmann scalar provides a long expression (which we shall not include here) which diverges at $r=0$ and all the $r_n$ above. As a result, the infinite coordinate singularities represent actual curvature singulatities.

What is worse, only the curvature singularity at $r=0$ can be out of reach for any causal observer, and only in very particular circumstances, as we shall see. To obtain this result, we first recall that $\Delta\lambda(r_\mathrm{ini}\rightarrow r_\mathrm{fin})$ can always be expressed in terms of the primitive \eqref{lambda primitive}, evaluated at some particular values of $r$.\footnote{Note, however, that expression \eqref{lambda difference} does not hold for arbitrary $r_\mathrm{ini}$, $r_\mathrm{fin}$ in this case. The reason is that, for Class 2 solutions with $\LCR_0>0$, the integrand in \eqref{lambda primitive} is a cosine, which changes sign periodically.} For solutions of the form \eqref{Class 2.1}, it turns out that
\begin{equation} \label{lambda Class 2.1}
    \lambda(r)=|C|\left[\sin D\,\Si\left(\sqrt{\dfrac{\LCR_0}{2}}r\right)+\cos D\, \Ci\left(\sqrt{\dfrac{\LCR_0}{2}}r\right)\right],
\end{equation}
where $\Si$ and $\Ci$ are the so-called sine integral and cosine integral functions, respectively:
\begin{gather}
    \Si(z)=\int_0^z \dfrac{\dif x}{x}\,\sin x, \label{Si} \\
    \Ci(z)=\gamma+\ln z+\int_0^z \dfrac{\dif x}{x}\,(\cos x-1), \label{Ci}
\end{gather}
where $\gamma$ is the Euler-Mascheroni constant. The sine integral function is regular for all $r$, while the cosine integral only blows up (to negative infinity) when $r=0$. As a result, $\lambda(r)$ solely becomes infinite at $r=0$ provided that $\cos D\neq 0$. There are, thus, two different scenarios:
\begin{itemize}
    \item If $\cos D\neq 0$, then $\Delta\lambda(r_\mathrm{ini}\rightarrow r_\mathrm{fin})$ blows up if and only if either $r_\mathrm{ini}$ or $r_\mathrm{fin}$ is equal to zero, and thus `only' the infinite number of curvature singularities located at $r_n$ ---with $r_n$ given by \eqref{r_n}--- are accessible for radially falling causal observers.
    \item If $\cos D=0$, then $\Delta\lambda(r_\mathrm{ini}\rightarrow r_\mathrm{fin})$ remains finite regardless of the values of $r_\mathrm{ini}$ and $r_\mathrm{fin}$. As a result, all curvature singularities can be reached in finite time by causal observers.
\end{itemize}
In either case, causally-propagating particles in Class 2 spacetimes with $\LCR_0>0$ may encounter an infinite number of curvature singularities in finite time, and thus we can immediately assert that these solutions are physically unjustifiable just for this reason. Other pathological properties of Class 2 solutions with $\LCR_0>0$ include the existence of infinite would-be Killing horizons located at each of the $r_n$ given by \eqref{r_n}---which cannot be true Killing horizons since their normal remains timelike for all $r$, as in Class 1 solutions with $C<0$---, as well as the inherent instability of $f(\LCR)$-exclusive constant-curvature solutions with $R_0\neq 0$ demonstrated in Result \ref{Result: R0 instability}.

\subsubsection{Class 2 solutions with $\LCR_0<0$}

Setting $B(r)=1$ and solving equation $\LCR=\LCR_0<0$ for $A(r)$, one obtains instead another set of novel solutions, characterised by
\begin{equation} \label{Class 2.2}
    \dif s^2=\left(\dfrac{C}{r}\right)^2\cosh^2\left(\sqrt{\dfrac{|\LCR_0|}{2}}\,r+D\right)\dif t^2-\dif r^2-r^2\dif\Omega^2,
\end{equation}
where $C\neq 0$ and $D$ are once again integration constants. Since the hyperbolic cosine never vanishes, \eqref{Class 2.2} only possesses one coordinate singularity, which is located at the origin $r=0$ of the spherical coordinate system. As in the $\LCR_0>0$ case, this coordinate singularity turns out to be a curvature singularity. The Kretschmann scalar for Class 2 solutions with $\LCR_0<0$ is given again by a convoluted expression (which we shall not include here); careful examination of such expression reveals that there are no more curvature singularities apart from the one at $r=0$.

Once more, we turn to determine whether the central curvature singularity is accessible to causal observers in spacetimes of the form \eqref{Class 2.2}. In this particular case, primitive \eqref{lambda primitive} might be readily expressed in terms of special functions, namely the hyperbolic sine integral (Shi) and the hyperbolic cosine integral (Chi), as\footnote{Notice that the integrand leading to \eqref{lambda Class 2.2} is a hyperbolic cosine, which is always positive. Thus, for Class 2 solutions with $\LCR_0<0$, identity \eqref{lambda difference} holds for whatsoever $r_\mathrm{ini}$, $r_\mathrm{fin}$.}
\begin{eqnarray} \label{lambda Class 2.2}
    \lambda(r)&=&|C|\left[\sinh D\,\Shi\left(\sqrt{\dfrac{|\LCR_0|}{2}}r\right)\right. \nonumber \\
    &&+\,\left.\cosh D\, \Chi\left(\sqrt{\dfrac{|\LCR_0|}{2}}r\right)\right].
\end{eqnarray}
Said special functions are defined as
\begin{gather}
    \Shi(z)=\int_0^z \dfrac{\dif x}{x}\,\sinh x, \\
    \Chi(z)=\gamma+\ln z-\int_0^z \dfrac{\dif x}{x}\,(\cosh x-1),
\end{gather}
in analogy with their trigonometric counterparts \eqref{Si} and \eqref{Ci}. Moreover, Shi, as Si, is regular for all $r$, while Chi, as Ci, diverges at $r=0$. Hence, $\lambda(r)$---and thus $\Delta\lambda(r_\mathrm{ini}\rightarrow r_\mathrm{fin})$---only diverges if either $r_\mathrm{ini}$ or $r_\mathrm{fin}$ is zero (notice that, this time, the pre-factor accompanying Chi in \eqref{lambda Class 2.2}, which is $\cosh D$, does not vanish for any $D$). As a result, the only singularity in \eqref{Class 2.2} is out of reach of radially infalling photons, and for this reason causal observers cannot encounter it in finite time.

The metric components of Class 2 solutions with $\LCR_0>0$ are always positive, so these metrics contain no apparent or Killing horizons. As a result, they describe the spacetime outside the singularity at $r=0$. As this singularity cannot interfere in finite time with causal observers, spacetimes of the form \eqref{Class 2.2} are much more physically well-founded than their counterparts \eqref{Class 2.1} with $\LCR_0<0$. Nonetheless, we insist that Class 2 spacetimes with $\LCR_0\neq 0$ are automatically unstable, as per Result \ref{Result: R0 instability}.

\subsubsection{Class 2 solutions with $R_0=0$}

The only Class 2 solutions which can be stable (in accordance with Result \ref{Result: R0=0 metastability}) are the ones which result from taking the limit of \eqref{Class 2.1} or \eqref{Class 2.2} as $\LCR_0\rightarrow 0$. It turns out that those solutions have a line element given by expression \eqref{Class 2.0}. This is the reason why we have grouped this solutions within Class 2 and not Class 1.

Line element \eqref{Class 2.0} appears to describe a wormhole with throat at $r=0$ upon extension of coordinate $r$ to negative values \cite{Calza:2018ohl}. This would be a remarkable result, since such a wormhole would not require negative energies to form; in fact, it would be a \textit{vacuum} solution of the equations of motion.

However, Class 2 solutions with $\LCR_0=0$ cannot describe wormholes since (i) they contain a naked singularity at $r=0$, and (ii) light rays cannot arrive at the locus of such singularity in finite affine parameter by following radial null geodesics. On the one hand, feature (ii) implies that the singularity is \textit{protected}; however, it also entails that, even if there were no curvature singularity, light rays would never be able cross the wrongly-identified wormhole throat in finite time.

The previous assertions (i) and (ii) are easy to verify. Indeed, the Kretschmann scalar corresponding to line element \eqref{Class 2.0} evaluates to
\begin{equation}
    \LCKre=\dfrac{24}{r^4},
\end{equation}
which diverges at $r=0$ independently of the value of $C$, as expected.\footnote{Notice that constant $C$ in line element \eqref{Class 2.0} can always be removed by redefining $C^2\,\dif t^2\rightarrow\dif t^2$; however, as we have done before for the other Class 2 solutions, we have explicitly included it in the metric it for dimensional reasons.} Using \eqref{Delta lambda}, we find that
\begin{equation}
    \Delta\lambda(r_\mathrm{ini}\rightarrow r_\mathrm{fin})=|C|\,\ln\left|\dfrac{r_\mathrm{fin}}{r_\mathrm{ini}}\right|,
\end{equation}
which only diverges when either $r_\mathrm{ini}$ or $r_\mathrm{fin}$ equals zero. Therefore, one cannot conclude that these solutions describe a traversable wormhole. In reality, what they model is two parallel and causally disconnected universes which are separated by an unreachable curvature singularity at the origin.

Apart from the central singularity, Class 2 solutions with $R_0=0$ do not have any other remarkable features; in particular, they have no apparent or Killing horizons.

\subsection{Solutions of Class 3} \label{Section: Class 3}

Class 3 solutions, discovered in Reference \cite{Calza:2018ohl}, can be obtained by solving equation $\LCR=\LCR_0$---with $R_0$ given, as always, by expression \eqref{static spherically symmetric Ricci scalar}---for $B(r)$, under the simple assumption that $A(r)=1$. Class 3 comprises a family of Kottler\footnote{The Kottler spacetime is also known as Schwarzschild-de Sitter or Schwarzschild-Anti de Sitter, depending on whether $\LCR_0$ is positive or negative, respectively.} lookalikes:
\begin{equation} \label{Class 3}
    \dif s^2=\dif t^2-\left(1-\dfrac{2GM}{r}-\dfrac{\LCR_0}{6}r^2\right)^{-1}\dif r^2-r^2\dif\Omega^2,
\end{equation}
where $M$ is a free parameter with units of mass, which can be (in principle) either positive or negative. Class 3 solutions with constant curvature $\LCR_0\neq 0$ are generically unstable as per Result \ref{Result: R0 instability}; Class 3 solutions with $\LCR_0=0$ could be at least metastable should the hypotheses of Result \ref{Result: R0 instability} hold.

Even thought line element \eqref{Class 3} is reminiscent of the Kottler metric, there are two notable differences between them. First, unlike the latter, the former have an anomalous redshift function distinct from unity; specifically,
\begin{equation} \label{Class 3 redshift}
    \Phi(r)=-\dfrac{1}{2}\ln\left(1-\dfrac{2GM}{r}-\dfrac{\LCR_0}{6}r^2\right).
\end{equation}
This modified-gravity-induced redshift function \eqref{Class 3 redshift} allows \eqref{Class 3} to have $g_{tt}=1$ instead of $g_{tt}g_{rr}=-1$. Second, the `cosmological-constant term' of Class 3 solutions, $\LCR_0 r^2/6$, is exactly half of the one present in Kottler spacetime, namely $\LCR_0 r^2/12$. As a result, Class 3 solutions have MSH mass
\begin{equation}
    M_\mathrm{MSH}(r)=M+\dfrac{\LCR_0}{12G}r^3.
\end{equation}
Notice that, while Class 3 solutions might harbour apparent horizons, they do not exhibit any Killing horizon corresponding to the generator of time translations $\partial_t$, since its norm remains equal to unity throughout all spacetime, as per \eqref{Killing norm} and \eqref{Class 3}.

While all members of this family were originally thought to describe traversable wormholes \cite{Calza:2018ohl}, we shall see that they might exhibit several unphysical properties depending on the values of $M$ and $\LCR_0$. We shall sort Class 3 solutions into sub-classes according to the signs of their free parameters $M$ and $\LCR_0$. We shall also consider the simple cases $M=0$, $\LCR_0\neq 0$ and $M\neq 0$ and $\LCR_0=0$ separately.

\subsubsection{Class 3 solutions with $M=0$ and $\LCR_0>0$}

In the simple case in which parameter $M$ vanishes and $R_0$ is positive, the metric reduces to a de Sitter lookalike:
\begin{equation} \label{Class 3 M = 0 R0 > 0}
    \dif s^2=\dif t^2-\left(1-\dfrac{\LCR_0}{6}r^2\right)^{-1}\dif r^2-r^2\dif\Omega^2,
\end{equation}
Class 3 solutions with $M=0$ have no curvature singularities, since their Kretschmann scalar, $\LCKre=\LCR_0^2/3$, remains constant for all $r$. Nonetheless, direct inspection of line element \eqref{Class 3 M = 0 R0 > 0} reveals that these spacetimes suffer from an evident physical pathology: if $r>r_\mathrm{ah}$, where
\begin{equation}
    r_\mathrm{ah}=\sqrt{\dfrac{6}{\LCR_0}},
\end{equation}
then the metric has two time coordinates, $t$ and $r$, since $g_{rr}$ becomes negative. What is more, the region with unphysical metric signature can be reached by causal observers in finite time. When evaluated for spacetime \eqref{Class 3 M = 0 R0 > 0} and $r_\mathrm{ini}$, $r_\mathrm{fin}\leq r_\mathrm{ah}$, integral \eqref{Delta lambda} can be computed using \eqref{lambda difference}, with primitive \eqref{lambda primitive} being given by
\begin{equation}
    \lambda(r)=\arcsin\left(\sqrt{\dfrac{\LCR_0}{6}}r\right).
\end{equation}
It is evident from this expression that $\Delta\lambda(r_\mathrm{ini}\rightarrow r_\mathrm{fin})$ remains finite when either $r_\mathrm{ini}$ or $r_\mathrm{fin}$ is equal to $r_\mathrm{ah}$.

Before closing this Section, it is interesting to emphasise that $r_\mathrm{ah}$ corresponds to an apparent horizon of \eqref{Class 3 M = 0 R0 > 0}, being a zero of $g^{rr}=1/B(r)$.

\subsubsection{Class 3 solutions with $M=0$ and $\LCR_0<0$}

One way to cure the pathology in the metric signature exhibited by \eqref{Class 3} is to change the sign of the constant curvature $R_0$. In such case, one obtains a line element given by
\begin{equation} \label{Class 3 M = 0 R0 < 0}
    \dif s^2=\dif t^2-\left(1+\dfrac{\LCR_0}{6}r^2\right)^{-1}\dif r^2-r^2\dif\Omega^2,
\end{equation}
whose components remain non-negative for all $r$, and has neither horizons nor curvature singularities.\footnote{Notice that, as in the previous case, the Kretschmann scalar is constant and equal to $\LCKre=R_0^2/3$. As a result, the usual coordinate singularity at $r=0$ is not an actual curvature singularity, but an artefact caused by the choice of `areal-radius' coordinates.} In fact, \eqref{Class 3 M = 0 R0 < 0} is the only Class 3 solution which truly describes a traversable wormhole centred at the origin, since the metric can be effortlessly extended to negative values of $r$. The remarkable fact that a traversable wormhole can exist as a vacuum solution of an infinite number of $f(\LCR)$ models is downplayed by the fact that such wormhole is generically unstable, as per Result \ref{Result: R0 instability}.

\subsubsection{Class 3 solutions with $M>0$ and $\LCR_0=0$}

In this case, the metric is a Schwarzschild lookalike:
\begin{equation} \label{Class 3 M > 0 R0 = 0}
    \dif s^2=\dif t^2-\left(1-\dfrac{2GM}{r}\right)^{-1}\dif r^2-r^2\dif\Omega^2.
\end{equation}
The metric signature changes to $(+,+,-,-)$ if $r<2GM$. Within this region also lies a curvature singularity, located at the origin $r=0$, as revealed by the form of the Kretschmann scalar, which is
\begin{equation} \label{Kretschmann Class 3 R0 = 0}
    \LCKre=\dfrac{24G^2M^2}{r^6}.
\end{equation}
Exactly as in the $M=0$, $R_0>0$ case, the unphysical region can be reached by radially infalling photons for a finite value of the affine parameter. This is because, for spacetime \eqref{Class 3 M > 0 R0 = 0}, integral \eqref{Delta lambda} is of the form \eqref{lambda difference}, where primitive \eqref{lambda primitive} is now given by
\begin{multline}
    \lambda(r)=\sqrt{r(r-2GM)} \\
    +GM\ln\left[r-GM+\sqrt{r(r-2GM)}\right].
\end{multline}
As we one may readily infer from this expression, $\lambda(r)$ remains finite and positive for all $r_\mathrm{ini}$ and $r_\mathrm{fin}$ equal or larger than $2GM$.\footnote{Notice that $\lambda(r)$ becomes complex for $r<2GM$, in yet another example of the pathological character of this sub-class of spacetimes.} In particular,
\begin{equation}
    \lambda(2GM)=GM\ln(GM).
\end{equation}
Hence, we conclude that line element \eqref{Class 3 M > 0 R0 = 0} is physically unsatisfactory, as causal observers can access the region with unphysical metric signature in finite time.

\subsubsection{Class 3 solutions with $M<0$ and $\LCR_0=0$}

A change in the sign of the mass parameter turns line element \eqref{Class 3 M = 0 R0 > 0} into
\begin{equation} \label{Class 3 M < 0 R0 = 0}
    \dif s^2=\dif t^2-\left(1+\dfrac{2G|M|}{r}\right)^{-1}\dif r^2-r^2\dif\Omega^2.
\end{equation}
As a result, $g_{rr}$ remains positive for all $r>0$, and thus the metric signature remains physical throughout all spacetime. Nonetheless, metric \eqref{Class 3 M < 0 R0 = 0} remains pathological, since the singularity at the origin is naked and reachable in finite proper time by causal observers. In this case, the Kretschmann scalar is still given by \eqref{Kretschmann Class 3 R0 = 0}, and thus $r=0$ is a true curvature singularity of \eqref{Class 3 M < 0 R0 = 0}. Furthermore, the affine-parameter interval between any two radii is given again by \eqref{lambda difference}, but with primitive $\lambda(r)$ now being
\begin{multline}
    \lambda(r)=\sqrt{r(r+2G|M|)} \\
    -G|M|\ln\left[r+G|M|+\sqrt{r(r+2G|M|)}\right].
\end{multline}
This quantity remains finite for all $r$, and in particular
\begin{equation}
    \lambda(0)=-G|M|\ln(G|M|);
\end{equation}
hence, the central singularity is causally connected to the rest of the spacetime. From this it is clear that, just as its positive-mass counterpart \eqref{Class 3 M > 0 R0 = 0}, metric \eqref{Class 3 M < 0 R0 = 0} is not well-founded from a physical point of view.

\subsubsection{Class 3 solutions with $M<0$ and $\LCR_0<0$}

When both the mass scale $M$ and the constant curvature $\LCR_0$ in \eqref{Class 3} are negative, the line element becomes 
\begin{equation} \label{Class 3 M < 0 R0 < 0}
    \dif s^2=\dif t^2-\left(1+\dfrac{2G|M|}{r}+\dfrac{|\LCR_0|}{6}r^2\right)^{-1}\dif r^2-r^2\dif\Omega^2.
\end{equation}
Thus, in this case, $g_{rr}$ remains positive for all positive radii, and the metric signature remains physical through all spacetime. However, the metric harbours a curvature singularity at $r=0$, where the Kretschmann scalar,
\begin{equation} \label{Class 3 same sign Kretschmann}
    \LCKre=\dfrac{24 G^2 M^2}{r^6}+\dfrac{\LCR_0^2}{3},
\end{equation}
diverges. Radially infalling photons can have access this singularity in finite time. This is because
\begin{eqnarray}
    \Delta\lambda(r_\mathrm{ini}\rightarrow 0)&=&\left|\int_{r_\mathrm{ini}}^0 \dif r\,\left(1+\dfrac{2G|M|}{r}+\dfrac{|\LCR_0|}{6}r^2\right)^{-1/2}\right| \nonumber \\
    &<&\left|\int_{r_\mathrm{ini}}^0 \dif r\,\right|=|r_\mathrm{ini}|<\infty
\end{eqnarray}
for any $r_\mathrm{ini}<\infty$. Therefore, this subclass is also pathological.

\subsubsection{Class 3 spacetimes with $M>0$ and $\LCR_0<0$}

In this case, the metric is
\begin{equation} \label{Class 3 M > 0 R0 < 0}
    \dif s^2=\dif t^2-\left(1-\dfrac{2GM}{r}+\dfrac{|\LCR_0|}{6}r^2\right)^{-1}\dif r^2-r^2\dif\Omega^2.
\end{equation}
Apart from the coordinate singularity at the origin, \eqref{Class 3 M > 0 R0 < 0} also has a coordinate singularity whenever there is an apparent horizon, i.e.~when $g^{rr}$ vanishes. This occurs only at
\begin{equation} \label{Apparent horizon Class 3 M > 0 R0 < 0}
    r_\mathrm{ah}=\sqrt{\dfrac{8}{|\LCR_0|}}\sinh\left[\dfrac{1}{3}\arcsinh\left(3GM\sqrt{\dfrac{|\LCR_0|}{2}}\right)\right],
\end{equation}
What is more, $g^{rr}$ is a strictly increasing function of $r$ for $r>0$. The monotonocity of $g^{rr}$, together with the fact that \eqref{Apparent horizon Class 3 M > 0 R0 < 0} is its only root, guarantees that $g^{rr}<0$ for $r<r_\mathrm{ah}$, and thus the metric signature changes inside the apparent horizon.

\subsubsection{Class 3 spacetimes with $M<0$ and $\LCR_0>0$}

This case is completely analogous to the previous one. The metric now reads
\begin{equation} \label{Class 3 M < 0 R0 > 0}
    \dif s^2=\dif t^2-\left(1+\dfrac{2G|M|}{r}-\dfrac{\LCR_0}{6}r^2\right)^{-1}\dif r^2-r^2\dif\Omega^2,
\end{equation}
and has a curvature singularity at $r=0$ and a coordinate singularity at
\begin{equation} \label{Apparent horizon Class 3 M < 0 R0 > 0}
    r_\mathrm{ah}=\sqrt{\dfrac{8}{\LCR_0}}\cosh\left[\dfrac{1}{3}\arccosh\left(3GM\sqrt{\dfrac{\LCR_0}{2}}\right)\right],
\end{equation}
which is also the only zero of $g^{rr}$, i.e.~an apparent horizon. Because $g^{rr}$ is monotonically decreasing for all $r$, we find that the metric signature changes for $r>r_\mathrm{ah}$.

\subsubsection{Class 3 spacetimes with $M>0$ and $\LCR_0>0$}

Last but not least, we proceed to analyse the case in which the mass and the scalar curvature of the Kottler lookalike are both positive. The line element becomes
\begin{equation} \label{Class 3 M > 0 R0 > 0}
    \dif s^2=\dif t^2-\left(1-\dfrac{2GM}{r}-\dfrac{\LCR_0}{6}r^2\right)^{-1}\dif r^2-r^2\dif\Omega^2.
\end{equation}
After some computations, one might realise that:
\begin{itemize}
    \item If $3GM\sqrt{\LCR_0/2}>1$, then $g^{rr}$ does not vanish for any $r$.
    \item If $3GM\sqrt{\LCR_0/2}=1$, then $g^{rr}$ has a double zero (i.e.~a wormhole throat) at $r=3GM=\sqrt{2/\LCR_0}$, which is also a coordinate singularity.
    \item If $0<3GM\sqrt{\LCR_0/2}<1$, then $g^{rr}$ vanishes at
    \begin{gather}
        r_0=\sqrt{\dfrac{8}{\LCR_0}}\sin\psi_0, \label{Apparent horizon Class 3 M > 0 R0 > 0 r_0} \\
        r_1=\sqrt{\dfrac{2}{\LCR_0}}\left(\sqrt{3}\cos\psi_0-\sin\psi_0\right), \label{Apparent horizon Class 3 M > 0 R0 > 0 r_1}
    \end{gather}
    where
    \begin{equation}
        \psi_0\equiv\dfrac{1}{3}\arcsinh\left(3GM\sqrt{\dfrac{\LCR_0}{2}}\right).
    \end{equation}
    Both $r_0$ and $r_1$ are apparent horizons and coordinate singularities.
\end{itemize}
Furthermore, it is straightforward to show that $g^{rr}$ always has a maximum at $r_*=(6GM/\LCR_0)^{1/3}$, with $g^{rr}(r_*)=1-3GM\sqrt{\LCR_0/2}$. In consequence,
\begin{itemize}
    \item If $3GM\sqrt{\LCR_0/2}>1$, then $g^{rr}(r_*)<0$, and thus $g^{rr}<0$ (i.e.~the metric signature is unphysical) for all $r$, since $r_*$ is a global maximum. Thus, this sub-class of solutions is not acceptable from a physical point of view.
    \item If $3GM\sqrt{\LCR_0/2}=1$, then $g^{rr}(r_*)=0$, and again the metric signature is unphysical for all $r$, due to the fact that $r_*$ is a maximum. For this reason, this sub-class of solutions is also not acceptable from a physical point of view.
    \item If $0<3GM\sqrt{\LCR_0/2}<1$, then $g^{rr}(r_*)>0$. Thus, because $r_0$ and $r_1$---as given by \eqref{Apparent horizon Class 3 M > 0 R0 > 0 r_0} and \eqref{Apparent horizon Class 3 M > 0 R0 > 0 r_0}, respectively---are zeroes of $g^{rr}$ (with $r_0<r_1$), we have that $g^{rr}$ must become negative for $r<r_0$ and $r>r_1$. Therefore, the metric signature is physical only for $r_0<r<r_1$.
\end{itemize}
Notice that all Class 3 solutions with $M>0$ and $\LCR_0>0$ also harbour a curvature singularity at the origin, as their Kretschmann scalar is given by \eqref{Class 3 same sign Kretschmann}. However, this curvature singularity lies within the region with unphysical metric signature.

\section{Conclusions} \label{Section: Conclusions}

In this work we have performed an assessment of the physical viability of $\LCR_0$-degenerate $f(\LCR)$ models, as well as of their (infinitely many) constant-curvature solutions. As our results demonstrate, there are reasons to believe that these models cannot be successful in describing Nature, even when experimental uncertainties do not rule them out directly.

One of the first anomalies we have detected is that $(\LCR_0=0)$-degenerate $f(\LCR)$ models feature an strongly-coupled Minkowski background (Result \ref{Result: Non-propagation of gravity}), meaning that the expected massless and traceless graviton does not propagate on top of such a background. This result renders $(\LCR_0=0)$-dependent $f(\LCR)$ models incompatible with gravitational-wave observations. For instance, this Result applies to the so-called `power-of-GR' models $f(\LCR)\propto R^{1+\delta}$ with integer $\delta>0$. We have also found that some $(\LCR_0=0)$-degenerate $f(\LCR)$ models do not even propagate the scalar degree of freedom atop a Minkowski background (Result \ref{Result: Non-propagation of the scalaron}), and thus their linearised spectrum does not contain any polarisation modes whatsoever.

Another important weakness of $\LCR_0$-degenerate $f(R)$ models (either with $\LCR_0=0$ or $\LCR_0\neq 0$) is their apparent lack of predictive power, given the infinite degeneracy of their constant-curvature solutions and the triviality of the equations of motion \eqref{f(R) EOM} when evaluated for $\LCR_0$-degenerate-exclusive constant-curvature solutions. More precisely, there is a subset of all possible initial conditions for the metric (namely, requiring $\LCR=\LCR_0$) whose evolution is not determined by the equations of motion of $\LCR_0$-degenerate models, which hold automatically for those initial conditions.

In relation to the previous point, there are reasons to believe that novel $f(\LCR)$-exclusive constant-curvature solutions can be easily matched to each other in $\LCR_0$-degenerate models \cite{Casado-Turrion:2022xkl}. This result is actually more disturbing than it seems, given that one expects most of the (infinitely many) constant-curvature solutions to exhibit all sorts of physically undesirable properties. In particular, the Class 1 models presented in Section \ref{Section: Class 1}, which are not exempt from pathologies, are known to smoothly match with Minkowski spacetime, forming a vacuole-like solution. The fact that one may freely choose the radius where solutions \eqref{Class 1.1} matches the interior Minkowski spacetime raises the question of whether a degenerate $f(\LCR)$ model can actually predict the boundaries at which spacetime ceases to be described by one of its constant-curvature solution and makes way for another constant-curvature solution.

Regarding the $\LCR_0$-degenerate constant-curvature solutions themselves, we have performed a stability analysis of these solutions, finding that, in general terms, the constant-curvature solutions of $(\LCR_0\neq 0)$-degenerate models are all unstable (Result \ref{Result: R0 instability}). The constant-curvature solutions of $(\LCR_0=0)$-degenerate $f(\LCR)$ models can be metastable provided that function $f$ satisfies some constraints (Result \ref{Result: R0=0 metastability}). This can be a problem, however, since the conditions required for the stability of constant-curvature solutions may be incompatible with those guaranteeing the absence of other instabilities (such as the Dolgov-Kawasaki instability). Moreover, the latter result also implies that constant-curvature solutions exhibiting pathological features can be stable.

In order to exemplify the kind of pathologies one may find in $f(\LCR)$-exclusive constant-curvature solutions, we have chosen three representative classes of such spacetimes, so as to thoroughly analyse four key aspects determining their physical viability: coordinate and curvature singularities, regions in which the metric acquires an unphysical signature (i.e.~the metric determinant changes sign due to one spatial coordinate abruptly becoming timelike), and geodesic completeness (i.e.~whether causal observers can or cannot encounter any of the previous pathologies in finite proper time). As the results in Table \ref{tab:solutions} reveal, all the solutions considered in this work exhibit unphysical properties. These unsubstantiated traits suffice to conclude that these solutions are unlikely to exist in Nature.

Finally, we must stress that there are some issues we would like to study in more detail in future works. For example, we have not investigated the linearised spectrum of $(\LCR_0\neq 0)$-degenerate $f(R)$ theories, in which the natural background is no longer Minkowski spacetime, but (Anti-)de Sitter (depending on the sign of $\LCR_0$). Similarly, we have not performed a perturbative expansion around any of the novel, $f(\LCR)$-exclusive constant-curvature solutions, since we have only been concerned with obtaining the linearised spectrum of the theory far from any gravitational-wave source. To shed more light on these issues, a general analysis of strong-coupling instabilities in $f(\LCR)$ gravity theories is currently in preparation.

\section*{Acknowledgements}

The authors would like to thank Sebasti\'{a}n Bahamonde, Valentin Boyanov, Luis J.~Garay, Gerardo Garc\'{\i}a-Moreno, Alejandro Jim\'{e}nez-Cano and Francisco Jos\'{e} Maldonado Torralba for their insightful comments and discussions.
The authors acknowledge support from project PID2019-108655GB-I00 (Ministerio de Ciencia e Innovaci\'{o}n, Spain).
ACT is also supported by a Universidad Complutense de Madrid-Banco Santander predoctoral contract CT63/19-CT64/19, as well as a Univesidad Complutense short-term mobility grant EB25/22.
\'{A}dlCD acknowledges further support from 
South African NRF grants no.120390, reference: BSFP190416431035; no.120396, reference: CSRP190405427545; and project PID2021-122938NB-I00 (Ministerio de Ciencia e Innovaci\'{o}n, Spain) and BG20/00236 action (Ministerio de Universidades, Spain). 
ACT would like to express his most sincere gratitude to all the members of the Laboratory of Theoretical Physics of the University of Tartu, for their kind help and hospitality during his three-month research stay in Estonia, as well as for their interest in the present work, which was in the last stages of preparation during said stay.

\appendix

\section{Some caveats concerning the use of the Einstein frame in $\LCR_0$-dependent $f(\LCR)$ models} \label{Appendix: Caveats}

In this Appendix, we shall present some of the subtleties arising when using the Einstein-frame representation \eqref{Einstein-frame metric}--\eqref{scalaron potential} of $\LCR_0$-degenerate $f(\LCR)$ models, and in particular when dealing with their constant-curvature solutions and their stability.

The first caveat is that conformal transformation \eqref{Einstein-frame metric} leading to the Einstein frame becomes singular when evaluated on $f(\LCR)$-exclusive constant-curvature solutions, since the existence of those solutions requires $f'(\LCR_0)=0$. In consequence, the dynamical equivalence between the Einstein and Jordan frames appears to break down in this scenario. However, given that the transformation between frames becomes singular only if $\LCR=\LCR_0$ (i.e.~in a null-measure set of values of $\LCR$), one may argue that the transformation does not indeed fail provided that all the relevant physical quantities remain well-defined after taking the limit $\LCR\rightarrow\LCR_0$.\footnote{Under this scope, the fact that the conformal transformation \eqref{Einstein-frame metric} becomes singular at $\LCR=\LCR_0$ is simply a reflection of the fact that the scalaron \eqref{scalaron} tends to negative infinity as $\LCR\rightarrow\LCR_0$.} In particular, the scalaron potential naively evaluates to a $0/0$ indetermination for $\LCR_0$-degenerate constant-curvature solutions (as mentioned in Section \ref{Section: Stability of solutions}). Only in cases where said indetermination can be resolved we shall consider the Einstein-frame to be a valid representation of the $f(\LCR)$ dynamics.

A second crucial observation is that, in our stability analysis, we have treated the scalaron potential $V$ as a function of $\LCR$ instead of as a function of the scalaron $\phi$, the reason being that it is more enlightening to perturb the scalar curvature instead of the abstract scalaron. However, in doing so, one must check that the extrema of $V$ as a function of $\LCR$ coincide with those of $V$ seen as a function of $\phi$. This is indeed the case provided that $f''(\LCR)\neq 0$. More precisely, given that
\begin{equation}
    \dfrac{\dif V}{\dif\LCR}=\dfrac{\dif\phi}{\dif\LCR}\dfrac{\dif V}{\dif\phi}=\sqrt{\dfrac{3}{2\kappa}}\dfrac{f''(\LCR)}{f'(\LCR)}\dfrac{\dif V}{\dif\phi},
\end{equation}
the zeroes of $\dif V/\dif R$ coincide with those of $\dif V/\dif\phi$ if and only if $f''(\LCR)\neq 0$. Moreover, one also has that
\begin{equation}
    \dfrac{\dif^2 V}{\dif\LCR^2}=\dfrac{\dif^2\phi}{\dif\LCR^2}\dfrac{\dif V}{\dif\phi}+\left(\dfrac{\dif\phi}{\dif\LCR}\right)^2\dfrac{\dif^2 V}{\dif\phi^2},
\end{equation}
by virtue of which
\begin{equation}
    \sign\left[\left.\dfrac{\dif^2 V}{\dif\LCR^2}\right|_{\frac{\dif V}{\dif\phi}=0}\right]=\sign\left[\dfrac{\dif^2 V}{\dif\phi^2}\right],
\end{equation}
i.e.~the character of the extremum (maximum, minimum or saddle point) does not change when one regards the scalaron potential to be a function of $\LCR$ instead of $\phi$.

Finally, it is worth mentioning that
\begin{equation}
    \dfrac{\dif V}{\dif\LCR}=\dfrac{f''(\LCR)}{2\kappa}\dfrac{2f(\LCR)-f'(\LCR) \LCR}{f'^3(\LCR)}.
\end{equation}
Thus, a constant-curvature solution with $\LCR=\LCR_0$ will extremise the scalaron potential---with respect to both $\LCR$ and $\phi$---provided that (i) $f''(\LCR_0)\neq 0$, (ii) $f'(\LCR_0)\neq 0$ and (iii) the trace \eqref{reduced f(R) EOM trace} of the equations of motion holds. If $f'(\LCR_0)=0$, i.e.~in $\LCR_0$-degenerate models, the stability analysis is more complicated and must be performed as done in Section \ref{Section: Stability of solutions}, as the potential and its derivatives might not be well defined in the limit $\LCR\rightarrow\LCR_0$. Moreover, when $f'(\LCR_0)=0$, mere compliance with equation \eqref{reduced f(R) EOM trace}---which now holds automatically---does not guarantee that $\LCR_0$-degenerate solutions extremise the potential. Results \ref{Result: R0 instability} and \ref{Result: R0=0 metastability} provide the conditions under which $\LCR_0$-degenerate constant-curvature solutions are not stable.

In order to shed more light on these issues, in Appendix \ref{Appendix: A simple model} we shall consider a particular instance of $f(\LCR)$ model hosting both unstable degenerate constant-curvature solutions and stable non-degenerate constant-curvature solutions.

\section{A simple illustrative model} \label{Appendix: A simple model}

For illustrative purposes, let us consider the simple, one-parameter model
\begin{equation} \label{simple model}
    f(\LCR)=\LCR_*-\LCR+\LCR\ln\left(\dfrac{\LCR}{\LCR_*}\right),
\end{equation}
where $\LCR_*>0$ is a constant with units of inverse length squared. This model harbours constant-curvature solutions for two different values of $\LCR$, namely:
\begin{itemize}
    \item $\LCR_0=\LCR_*$. Since $f(\LCR_*)=0$ and $f'(\LCR_*)=0$, all spacetimes with $\LCR_0=\LCR_*$ trivially solve the equations of motion associated to \eqref{simple model}, i.e.~the model is $\LCR_*$-degenerate.
    \item $\LCR_0=\eta\LCR_*$, with $\eta=4.92155...$ being the non-trivial solution of $2-2\eta+\eta\ln\eta=0$ ($\eta=1$ is the trivial solution of this transcendental equation). In this case, $f(\eta\LCR_*)\neq 0$ and $f'(\eta\LCR_*)\neq 0$, but the trace of the equations of motion \eqref{reduced f(R) EOM trace} holds. As a result, the $f(\LCR)$ equations of motion \eqref{f(R) EOM} reduce to the Einstein equations with cosmological constant \eqref{reduced f(R) EOM}, and model \eqref{simple model} admits the same constant-curvature solutions as $\text{GR}+\Lambda$, provided $\Lambda=\eta\LCR_*/4$, without being $(\eta\LCR_*)$-degenerate.
\end{itemize}
Due to Result \ref{Result: R0 instability}, the constant-curvature solutions of \eqref{simple model} having $\LCR_0=\LCR_*$ are all unstable. However, the non-degenerate constant-curvature solutions with $\LCR_0=\eta\LCR_*$ are all stable. Indeed, it is straightforward to check that (i) $f''(\eta\LCR_*)\neq 0$, (ii) $f'(\eta\LCR_*)\neq 0$ and (iii) equation \eqref{reduced f(R) EOM trace} is satisfied for constant-curvature solutions with $\LCR_0=\eta\LCR_*$. Thus, as per the discussion in Appendix \ref{Appendix: Caveats}, non-degenerate solutions having $\LCR_0=\eta\LCR_*$ extremise the scalaron potential, regardless of whether it is understood to be a function of $\LCR$ or of $\phi$. Moreover,
\begin{equation}
    \left.\dfrac{\dif V}{\dif\LCR}\right|_{\LCR=\eta\LCR_*}=\dfrac{\eta}{8\LCR_*}\dfrac{\eta-2}{(\eta-1)^3}>0,
\end{equation}
given that $\LCR_*>0$ and $\eta>2$. Thus, $\LCR_0=\eta\LCR_*$ is a minimum of the scalaron potential. It can be shown that there are no other extrema in the region $\LCR>\LCR_0$, and given that $V(\LCR)\rightarrow+\infty$ as $\LCR\rightarrow\LCR_0^+$, we conclude that non-degenerate constant-curvature solutions of model \eqref{simple model} having $\LCR_0=\eta\LCR_*$ are stable.

\section{Quantities and formulae of interest for the physical characterisation of solutions} \label{Appendix: Characterisation}

Given the extraordinary amount of symmetry exhibited by static, spherically symmetric spacetimes of the form \eqref{static spherically symmetric}, the study of all the points listed at the beginning of Section \ref{section:constant-curvature solutions} simplifies considerably. For any given solution, one essentially needs to determine the values of $r$ where functions $A(r)$ and $B(r)$ either vanish or become infinite, and which of these points or regions are pathological, as explained in what follows. As stated in the bulk of the text, this Appendix is also meant to further clarify our nomenclature and symbol conventions.

\subsection{Apparent and Killing horizons}

It can be shown that, when a static spherically symmetric spacetime is expressed in the Abreu-Nielsen-Visser gauge \eqref{Abreu-Nielsen-Visser}---equivalently, in areal-radius coordinates \eqref{static spherically symmetric}---, its apparent horizons are located at the \textit{simple} roots $r_\mathrm{ah}$ of the algebraic equation $g^{rr}(r_\mathrm{ah})=0$.\footnote{In addition, spacetimes \eqref{Abreu-Nielsen-Visser} possess a wormhole throat wherever $g^{rr}$ has a \textit{double} root.} In other words, \eqref{static spherically symmetric} has an apparent horizon at $r=r_\mathrm{ah}$ when $B(r_\mathrm{ah})\rightarrow\infty$ or, equivalently, if
\begin{equation} 
\label{apparent horizon ANV gauge}
    r_\mathrm{ah}=2G M_\mathrm{MSH}(r_\mathrm{ah}).
\end{equation}

On the other hand, $\xi=\partial/\partial t$ is a Killing vector of every line element of the form \eqref{static spherically symmetric}, since they are all static. As a result, these spacetimes will harbour a Killing horizon provided that Killing vector $\xi$ becomes null in some region of spacetime. In areal-radius coordinates $(t,r,\theta,\varphi)$, $\xi^\mu=\delta^\mu_{\hphantom{\mu}t}$, so the norm of $\xi$ is given by
\begin{equation} \label{Killing norm}
    g_{\mu\nu}\xi^\mu\xi^\nu=g_{tt}=A(r).
\end{equation}
We then conclude that spacetimes of the form \eqref{static spherically symmetric} will host a Killing horizon at $r=r_\mathrm{Kh}$ provided that $r_\mathrm{Kh}$ is a root of
\begin{equation} 
\label{Killing horizon static spherically symmetric areal}
    A(r_\mathrm{Kh})=0.
\end{equation}
Expressions \eqref{apparent horizon ANV gauge} and \eqref{Killing horizon static spherically symmetric areal} imply that line element \eqref{static spherically symmetric} exhibits coordinate singularities at the locations of the apparent and Killing horizons, respectively. However, we must stress that the aforementioned conditions \eqref{apparent horizon ANV gauge} and \eqref{Killing horizon static spherically symmetric areal} are obtained by computing scalar quantities, which should remain invariant even when they are expressed in singular coordinates, such as the areal radius coordinates. For the purpose of our analysis, this level of rigour shall be sufficient; we are nonetheless aware that a more detailed and mathematically precise computation can be performed.

\subsection{Singularities and geodesic completeness}

The existence of coordinate singularities (points in which $A(r)$ and/or $B(r)$ become zero or infinite) may also point out the presence of curvature singularities. The existence of such curvature singularities, however, must be determined in a coordinate-invariant way; for example, we will compute the Kretschmann scalar
\begin{equation}
    \LCKre=R_{\mu\nu\rho\sigma}R^{\mu\nu\rho\sigma}
\end{equation}
for each solution, and then determine the values of $r$ where this quantity diverges.

The mere existence of curvature singularities is not a sign of unphysical dynamics \textit{per se}, even though they represent points or regions in which tidal forces become infinite. Should these singularities be unreachable in finite affine parameter for causal observers, then none of such observers would experience infinite tidal forces at any point along their world-lines.

In order to compute whether a photon can reach a singularity for a finite value of the affine parameter, we will analyse the corresponding geodesic equation. Given that all the spacetimes we are considering are spherically symmetric, we may always choose, without loss of generality, to perform all computations on the equatorial plane ($\theta=\pi/2$). As a result, the equation for the null geodesics of \eqref{static spherically symmetric} reduces to
\begin{equation} \label{static spherically symmetric null geodesic eq equatorial}
    \left(\dfrac{\dif r}{\dif\lambda}\right)^2=\dfrac{E^2}{A(r)B(r)}-\dfrac{L^2}{r^2B(r)}\,,
\end{equation}
where $\lambda$ is the affine parameter of the trajectory and $E$ and $L$ are the observer's conserved quantities, namely
\begin{gather}
    E=A(r)\,\dfrac{\dif t}{\dif\lambda}=\const\,, \label{geodesic energy} \\
    L=r^2\,\dfrac{\dif\varphi}{\dif\lambda}=\const
\end{gather}
From expression \eqref{geodesic energy}, it is evident that one can always rescale $\lambda$ in such a way that $E=1$. Accordingly, a photon with unit energy travelling from a given $r_\mathrm{ini}$ following a radial geodesic ($L=0$) takes the following variation of affine parameter $\lambda$ to reach any other value of the areal radius $r_\mathrm{fin}$:
\begin{equation} \label{Delta lambda}
    \Delta\lambda(r_\mathrm{ini}\rightarrow r_\mathrm{fin})=\left|\int_{r_\mathrm{ini}}^{r_\mathrm{fin}}\dif r\,\sqrt{A(r)B(r)}\right|\,.
\end{equation}
(Notice that an absolute value has been intentionally included in the previous expression in order to produce the same value of $\Delta\lambda(r_\mathrm{ini}\rightarrow r_\mathrm{fin})$ regardless of whether $r_\mathrm{ini}<r_\mathrm{fin}$ or $r_\mathrm{ini}>r_\mathrm{fin}$.) In most practical cases, the evaluation of \eqref{Delta lambda} will reduce to
\begin{equation} \label{lambda difference}
    \Delta\lambda(r_\mathrm{ini}\rightarrow r_\mathrm{fin})=|\lambda(r_\mathrm{fin})-\lambda(r_\mathrm{ini})|\,,
\end{equation}
where we have defined the primitive
\begin{equation} \label{lambda primitive}
    \lambda(r)\equiv\int\dif r\,\sqrt{A(r)B(r)}\,.
\end{equation}
However, it is evident that \eqref{lambda difference} does not hold when the integrand $\sqrt{A(r)B(r)}$ changes sign within the integration interval, as is the case with some of the spacetimes and intervals considered in this work.

\subsection{Regions in which the metric signature becomes unphysical}

Last but not least, the existence of zeroes of functions $A$ and $B$ can also lead to changes in the metric signature. For example, if $A$ becomes negative for some values of $r$, then the metric becomes Euclidean (i.e.~all coordinates become space-like). If, on the contrary, $B$ becomes negative, then there are two time coordinates. Both situations are clearly unphysical and should be avoided. In particular, we can use again equation \eqref{Delta lambda} to know whether such regions with unphysical metric signatures can be reached by causal observers in finite affine parameter.

\nocite{*}
\bibliography{bibliography}

\end{document}